\documentclass[jpc,twocolumn,datetime]{revtex4-1}  
\usepackage{amsfonts}
\usepackage{amsmath}
\usepackage{amssymb}
\usepackage{version}
\usepackage{cancel}
\usepackage{graphicx}%
\includeversion{New_connection}
\excludeversion{Old_connection}

\begin{document}
\noindent
Journal of Statistical Physics {\bf 193}, 12 (2026)\\
https://doi.org/10.1007/s10955-025-03553-3\\

\title{Linear Analysis of Stochastic Verlet-Type Integrators for Langevin Equations}
\author{Niels Gr{\o}nbech-Jensen}
\email{ngjensen@ucdavis.edu}
\affiliation{Department of Mechanical \& Aerospace Engineering
\\ Department of Mathematics\\ University of California, Davis, CA 95616, U.S.A.}

\begin{abstract}



\noindent
We provide an analytical framework for analyzing the quality of stochastic Verlet-type integrators for simulating the Langevin equation. Focusing only on basic objective measures, we consider the ability of an integrator to correctly simulate two characteristic configurational quantities of transport, a) diffusion on a flat surface and b) drift on a tilted planar surface, as well as c) statistical sampling of a harmonic potential. For any stochastic Verlet-type integrator expressed in its configurational form, we develop closed form expressions to directly assess these three most basic quantities as a function of the applied time step. The applicability of the analysis is exemplified through twelve representative integrators developed over the past five decades, and algorithm performance is conveniently visualized through the three characteristic measures for each integrator. The GJ set of integrators stands out as the only option for correctly simulating diffusion, drift, and Boltzmann distribution in linear systems, and we therefore suggest that this general method is the one best suited for high quality thermodynamic simulations of nonlinear and complex systems, including for relatively high time steps compared to simulations with other integrators.
\end{abstract}

\maketitle
\section{Introduction}
\label{sec:intro}
The prevailing numerical methods for time-dependent computational statistical mechanics and molecular dynamics are based on the simple and efficient Newton-St{\o}rmer-Verlet discretization of Newton's equation of motion. The reason is that this framework strikes a seemingly optimal balance between accuracy, efficiency, and simplicity in addition to exhibiting critical conservation properties that ensure consistency throughout long simulations. While this very simple temporal discretization had been introduced by Newton (see, e.g., Ref.~\cite{Toxsvaerd}), and later further explored by St{\o}rmer (see, e.g., Ref.~\cite{Stormer_1921}), the usefulness for modern simulations grew dramatically after Verlet's re-introduction of this configurational method in 1967 \cite{Verlet}, and the method has since then been commonly referred to as the Verlet method. The next few years produced augmented algorithms that possessed the same configurational evolution, but with the addition of central difference approximations to the corresponding velocity variable at either the time step of the configurational coordinate (on-site) \cite{Swope,Beeman} or at the half-step between two time steps of the coordinate \cite{Buneman,Hockney}, the former typically denoted the velocity-Verlet method and the latter the leap-frog method. A formulation that naturally produces both half-step and on-site velocities is shown in Ref.~\cite{Tuckerman}. Applications and implementations, especially those relating to molecular modeling, are detailed in, e.g., Refs.~\cite{AllenTildesley,Frenkel,Rapaport,Hoover_book,Leach}. Regardless of the formulations that include different velocity approximations, the underlying discrete-time configurational evolution is the same Verlet method \cite{Verlet}.

For Langevin modeling, where Newton's equation of motion is revised to satisfy the fluctuation-dissipation balance with an imaginary heat-bath, the Verlet method has similarly been revised to include friction and thermal noise contributions in discrete time. While the Verlet method for Newtonian deterministic evolution contains only the selected time step as a free parameter, incorporating friction and noise into the discrete-time Verlet framework can be done in a vast number of seemingly similar ways through different derivation strategies. This leads to a large number of stochastic numerical integrators (thermostats) with different properties, as exemplified by the descriptions in Refs.~\cite{SS,Ermak1980,Allen80,Allen82,Thalmann2007,vGB82,BBK,vGB88,GW97,Skeel2002,Ricci,VEC06,LM,GJ}, which contain the subset of integrators specifically analyzed in this work. Given the number of available integrators, which all converge to the same (correct) behavior in the small time step limit, but all diverge from each other as the time step is increased, valuable contributions to the literature have been the occasional publications that seek to compare algorithm performance of a few select integrators of interest (see, e.g., Refs.~\cite{Pastor_88,2GJ,Finkelstein_1,Tanygin2024}), thereby providing some insight to which integrator to choose. Such investigations often select a model problem from a complex system, such as a molecular dynamics ensemble, and use the results to draw conclusions. Comparisons are typically made for both configurational and kinetic coordinates, extracting relevant thermal statistics of potential and kinetic energies, diffusion constants obtained from both configurational and kinetic evolution, etc. Yet, the mutual dependence between configurational and kinetic coordinates that exists in continuous time does not exist in discrete time, as evidenced by, e.g., the Verlet algorithm mentioned above. Indeed, the same configurational evolution can be obtained with or without a velocity variable, and many different velocity variables can be defined to accompany any given configurational trajectory \cite{2GJ,GJ,GJ24}. Additionally, many, if not most, Langevin simulations seek to obtain configurational statistics, such as sampling distribution of phase space or transport in the form of either diffusion or drift. Thus, we submit that the core of an evaluation of an integrator is to first establish its configurational properties, and if those properties are desirable, then decide on the design of a velocity variable that can be consistently associated with the integrator in a manner in which, e.g., simulated kinetic and configurational temperatures coincide (see Refs.~\cite{2GJ,GJ24}).

Following the spirit of Ref.~\cite{Pastor_88}, the presumption of this work is that the quality of a numerical Langevin integrator applied to a nonlinear, complex problem starts with understanding its configurational features for simple, linear problems. The rationale for this approach is that 1) linear analysis can be done objectively for all stochastic Verlet-type integrators, as will be done here, 2) since a thermodynamic system seeks the free energy minimum of an energy surface, the linear component of a problem is often a significant contributor to the sampling, and 3) good algorithm performance for complex problems is expected to be correlated to good performance for simple, linear problems; or expressed in reverse, it is unlikely that an algorithm that is inaccurate for sampling the statistics of linear problems will be generally reliable for nonlinear ones.

It is the purpose of this paper to develop a general set of expressions to evaluate stochastic integrators. Within this framework we select a set of commonly discussed methods for configurational evaluation of their basic linear properties relevant for computational statistical mechanics, and thereby provide an objective, problem-independent sense of what one should expect from these integrators. To that end, for any stochastic Verlet thermostat expressed in its pure configurational form, we analytically derive closed expressions for the three most basic properties that are characteristic of the three kinds of linear forces; namely a) diffusion constant for an object on a flat potential, b) drift velocity of an object on a tilted planar potential, and c) the sampling distribution of the location of an object in a harmonic potential. These expressions provide a direct path for basic algorithm analysis, and they provide insight to how an algorithm must be structured in order to yield statistically accurate simulation results as a function of the applied time-step. Twelve representative stochastic Verlet-type integrators are selected and analyzed in chronological order. Based on the analytical expressions for the three statistical quantities of interest, we solidify precisely how a stochastic Verlet-type integrator must be designed to give correct statistical response for linear systems regardless of the time step.

\section{Background}
\label{sec:Background}
Each degree of freedom of the physical, continuous-time system of interest is modeled by the Langevin equation \cite{Langevin,Langevin_Eq} 
\begin{eqnarray}
m\ddot{r}+\alpha\dot{r} & = & f+\beta \; , \label{eq:Langevin}
\end{eqnarray}
for an object with mass $m$ and location (configurational coordinate) $r$. The mass is subject to a force $f=-\nabla E_p$, where $E_p(r)$ is a potential energy surface, and a linear friction force $-\alpha \dot{r}$ given by the damping coefficient $\alpha\ge0$. The associated thermal noise force, $\beta$, is given by the fluctuation-dissipation relationship \cite{Parisi},
\begin{subequations}
\begin{eqnarray}
\langle\beta(t)\rangle & = & 0 \\
\langle\beta(t)\beta(t^\prime)\rangle & = & 2\alpha\, k_BT\, \delta(t-t^\prime) \; , 
\end{eqnarray}
\label{eq:FD}\noindent
\end{subequations}
where $T$ is the temperature of the heat bath, $k_B$ is Boltzmann's constant, $\delta(t)$ is Dirac's delta function, and $\langle\cdot\rangle$ represents a statistical average.

Our interest in this work is the three key configurational results for linear systems \cite{Langevin_Eq}:

\noindent
{\bf a)} Diffusion constant $D_E$ for a flat potential ($f=0$)
\begin{subequations}
\begin{eqnarray}
D_E & = & \lim_{s\rightarrow\infty}\frac{\left\langle[r(t+s)-r(t)]^2\right\rangle}{2s} \; = \; \frac{k_BT}{\alpha}\; \; , \; \alpha>0 \, ,\nonumber \\ \label{eq:Diff_rr}
\end{eqnarray}\label{eq:Cont_time_quantities}\noindent
as defined by the configurational Einstein expression;

\noindent
{\bf b)} Drift velocity $v_d$ for a tilted, planar potential ($f={\rm const}$)
\begin{eqnarray}
v_d & = & \frac{\langle r(t+s)-r(t)\rangle}{s} \; = \; \frac{f}{\alpha}\; \; , \; \alpha>0\; , \;  s\neq0\, ;  \label{eq:Drift_r}
\end{eqnarray}
and

\noindent
{\bf c)} Configurational temperature $T_c$ for the harmonic potential, $E_p(r)=\frac{1}{2}\kappa r^2$ ($f=-\kappa r$),
\begin{eqnarray}
k_BT_c & = & 2\left\langle E_p\right\rangle \; = \; \kappa\left\langle r(t)r(t)\right\rangle \; = \; k_BT \; \; , \; \kappa>0 \, , \label{eq:Corr_rr}
\end{eqnarray}\label{eq:Basic_truths}\noindent
\end{subequations}
where the configurational temperature \cite{Landau,Rugh,Powles_05,Saw_23},
\begin{eqnarray}
T_c & = & \frac{1}{k_B}\frac{\left\langle(\partial E_p/\partial r)^2\right\rangle}{\left\langle\partial^2E_p/\partial r^2\right\rangle} \, , \label{eq:Configurational}
\end{eqnarray}
reduces to the relationships given in Eq.~(\ref{eq:Corr_rr}) for a Hooke's force, and where $\langle r(t)r(t)\rangle$ is the variance of the configurational sampling distribution of the potential $E_p(r)$.

The three relationships of Eq.~(\ref{eq:Cont_time_quantities}) are the basic benchmarks we will use to characterize discrete-time stochastic Verlet thermostats based on linear analysis.

\section{The Configurational Stochastic Verlet-Type Integrator}
\label{sec:Config}
As mentioned above, we will here focus exclusively on the statistical properties of the discrete-time {\it configurational} coordinate since many different definitions of accompanying discrete-time velocities with different properties can be formulated for any given configurational integrator \cite{GJ24}. Thus, the key to understanding the properties of an integrator lies in its configurational coordinate.

The rather broad class of Verlet-type stochastic integrators, where only one evaluation of the force field per time step is necessary, can be analyzed entirely without a velocity by writing the integrator in the general form,
\begin{eqnarray}
r^{n+1} & = & 2c_1r^n-c_2r^{n-1}+\frac{\Delta{t}^2}{m}(c_3f^n+c_4f^{n-1})\nonumber \\
&& +\frac{\Delta{t}}{2m}(c_5\beta_-^n+c_6\beta_+^n) \, , \label{eq:Vt_general}
\end{eqnarray}
where the discrete-time coordinates at time $t_n$ are $r^n$ and $f^n=f(r^n)$, and  the time step is $\Delta{t}=t_{n+1}-t_n$. The integrated fluctuations $\beta_\pm^n$ may be written,
\begin{subequations}
\begin{eqnarray}
c_5\beta_-^n & = & \int_{t_{n-1}}^{t_n}\psi(t-t_n)\,\beta(t)\, dt \label{eq:d_noise_-} \\
c_6\beta_+^n & = & \int_{t_n}^{t_{n+1}}\psi(t-t_n)\,\beta(t)\, dt\,,  \label{eq:d_noise_pm} 
\end{eqnarray}\label{eq:d_noise_both}\noindent
\end{subequations}
with $\psi(s)$ being a chosen weight function on the interval, $s\in[-\Delta{t},\Delta{t}]$, such that
\begin{subequations}
\begin{eqnarray}
\left\langle\beta_\pm^n\right\rangle & = & 0 \label{eq:d_noise_ave}\\
\left\langle\beta_\pm^n\beta_\pm^\ell\right\rangle & = & 2\,\alpha\Delta{t}\, k_BT \, \delta_{n,\ell} \label{eq:d_noise_auto}\\
\left\langle\beta_\pm^n\beta_\mp^\ell\right\rangle & = & 2\,\alpha\Delta{t}\, k_BT\, \zeta\, \delta_{n,\ell\mp1}\, , \label{eq:d_noise_cor}
\end{eqnarray} \label{eq:d_noise_all}\noindent
\end{subequations}\noindent
the correlation, $-1\le\zeta\le1$, being defined in Eq.~(\ref{eq:d_noise_cor}). Notice that $\zeta\ge0$ if $\psi(s)$ does not change its sign in $s\in[-\Delta{t},\Delta{t}]$. Notice also that while many different stochastic variables, $\beta(t)$, can satisfy Eq.~(\ref{eq:FD}), the corresponding discrete-time noise $\beta_\pm^n$ must be chosen from a Gaussian distribution as outlined in Ref.~\cite{Gauss_noise}, and this work therefore considers only Gaussian fluctuations. The unit-less functional parameters, $c_i$ and $\zeta$, depend only on $\gamma\Delta{t}$, where 
\begin{eqnarray}
\gamma & = & \frac{\alpha}{m} \, .
\end{eqnarray}
The temporal discretization of the differentials in Eq.~(\ref{eq:Vt_general}) requires that
\begin{eqnarray}
2c_1 & = & 1+c_2 \, , \label{eq:c1}
\end{eqnarray}
where $c_2$ is the one-time-step velocity attenuation parameter \cite{GJ}, for which stability necessitates $|c_2|\le1$ with the expected limit, $c_2\rightarrow1$ for ${\gamma\Delta{t}}\rightarrow0$. Stochastic Verlet-type integrators are defined by their expressions of $c_i$ and $\zeta$ as functions of $\gamma\Delta{t}$, and their algorithmic properties can therefore be entirely characterized from these parameters.

Expanding on Refs.~\cite{GJ,Josh_2020}, Appendices \ref{app:app_Diff}, \ref{app:app_Drift}, and \ref{app:app_Dist} derive, for linear systems as described above, the complete expressions of, respectively, {\it diffusion} [Eq.~(\ref{eq:DE_discrete})], {\it drift} [Eq.~(\ref{eq:Drift_discrete})], and the variance of the {\it configurational sampling} distribution [Eq.~(\ref{eq:Dist_discrete})] as functions of the parameters, $c_i$ and the correlation $\zeta$, in Eqs.~(\ref{eq:Vt_general}) and (\ref{eq:d_noise_all}).  The resulting discrete-time quantities, $\Gamma_{\rm diff}$, $\Gamma_{\rm drift}$, and $\Gamma_{\rm dist}$, are normalized to the corresponding correct continuous-time quantities given in Eq.~(\ref{eq:Basic_truths}). Clearly, the optimal value for these three normalized quantities is $\Gamma_{\rm diff}=\Gamma_{\rm drift}=\Gamma_{\rm dist}=1$. In contrast to Refs.~\cite{GJ,Josh_2020}, we here maintain the full dependence on each of the functional parameters in order to make the expressions applicable to the general set of stochastic Verlet-type integrators. Appendix~\ref{app:app_Stability} outlines the various stability conditions that apply to the general integrator in Eq.~(\ref{eq:Vt_general}), and Appendix~\ref{app:app_Exclusive} solidifies the only integrator parameter set that accomplishes the objective, $\Gamma_{\rm diff}=\Gamma_{\rm drift}=\Gamma_{\rm dist}=1$.

\section{Investigation of Specific Integrators}
\label{sec:Specific_Int}
With the derived key expressions, $\Gamma_{\rm diff}$ (from Eq.~(\ref{eq:DE_discrete})), $\Gamma_{\rm drift}$ (from Eq.~(\ref{eq:Drift_discrete})), and $\Gamma_{\rm dist}$ (from Eq.~(\ref{eq:Dist_discrete})), we now turn to methodically consider specific methods, evaluating their configurational properties based on the characteristic behavior of  $\Gamma_{\rm diff}$, $\Gamma_{\rm drift}$, and $\Gamma_{\rm dist}$ as a function of the reduced time step. Notice that $\Gamma_{\rm diff}$ and $\Gamma_{\rm drift}$ are functions of only $\gamma\Delta{t}$, while $\Gamma_{\rm dist}$ is a function of both $\gamma\Delta{t}$ and $\Omega_0\Delta{t}$, where $\Omega_0^2=\kappa/m$ is the natural frequency of the harmonic oscillator that is associated with $\Gamma_{\rm dist}$; see Appendix~\ref{app:app_Dist}. We consider the twelve selected integrators in chronological publication order, while we emphasize that any other stochastic Verlet-type integrator can be similarly analyzed by the derived expressions in Appendices \ref{app:app_Diff}, \ref{app:app_Drift}, and \ref{app:app_Dist}.

\subsection{{{SS$_{78}$}}}
\label{sec:sec_SS}

\begin{figure}[t]
\centering
\scalebox{0.585}{\centering \includegraphics[trim={3.10cm 13.75cm 1cm 5.0cm},clip]{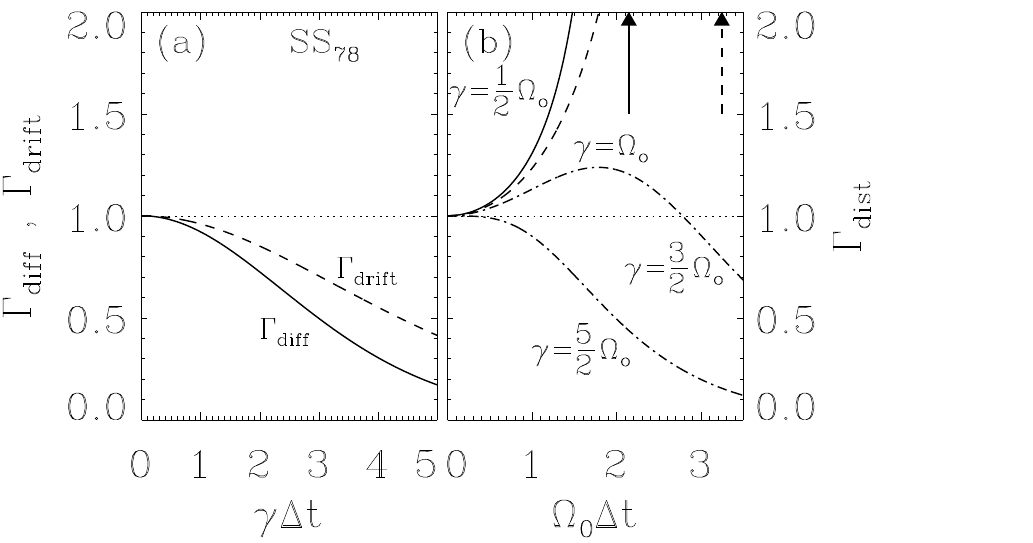}}
\caption{Normalized discrete-time diffusion [$\Gamma_{\rm diff}$, Eqs.~(\ref{eq:DE_discrete}) and (\ref{eq:SS_G_diff})], drift [$\Gamma_{\rm drift}$, Eqs.~(\ref{eq:Drift_discrete}) and (\ref{eq:SS_G_drift})], and Boltzmann configurational temperature [$\Gamma_{\rm dist}$, Eqs.~(\ref{eq:Dist_discrete}) and (\ref{eq:SS_G_dist})] for the SS$_{78}$ integrator of Ref.~\cite{SS}, as a function (a) of  $\gamma\Delta{t}$ for $\Gamma_{\rm diff}$ (solid) and $\Gamma_{\rm drift}$ (dashed), and (b) of $\Omega_0\Delta{t}$ for $\Gamma_{\rm dist}$, the latter for select values of normalized damping $\gamma/\Omega_0$ as indicated on the figure. Vertical arrows in (b) indicate the stability limit of the time step as given by Eq.~(\ref{eq:Stability_eq_R-}). Discrete-time quantities are normalized to the correct continuous-time quantities, Eq.~(\ref{eq:Basic_truths}).
}
\label{fig:fig_SS}
\end{figure}

The SS$_{78}$ integrator by Schneider and Stoll \cite{SS} originates in the configurational form of Eq.~(\ref{eq:Vt_general}) with the functional coefficients
\allowdisplaybreaks\begin{subequations}
\begin{eqnarray}
c_2 & = & 2c_1-1 \; = \; e^{-\gamma\Delta{t}} \label{eq:SS_c2}\\
c_3 & = & \sqrt{c_2} \; \rightarrow \; 1 \; \; {\rm for} \; \; \gamma\Delta{t}\rightarrow0 \label{eq:SS_c3} \\
c_4 & = & 0 \label{eq:SS_c4} \\
c_5 & = & 0 \label{eq:SS_c5} \\
c_6 & = & 2\sqrt{c_2}  \; \rightarrow \; 2 \; \; {\rm for} \; \; \gamma\Delta{t}\rightarrow0\label{eq:SS_c6} \\
\zeta & = & 0 \,. \label{eq:SS_zeta}
\end{eqnarray}\label{eq:SS_coeff}\noindent
\end{subequations}
Inserting these coefficients into Eqs.~(\ref{eq:DE_discrete}), (\ref{eq:Drift_discrete}), and (\ref{eq:Dist_discrete}) yields the normalized characteristic measures
\begin{subequations}
\begin{eqnarray}
\Gamma_{\rm diff} & = & \left(\frac{{\gamma\Delta{t}}}{1-c_2}\right)^2c_2
 \label{eq:SS_G_diff}\\
\Gamma_{\rm drift} & = & \frac{{\gamma\Delta{t}}}{1-c_2}\sqrt{c_2} 
 \label{eq:SS_G_drift}\\
\Gamma_{\rm dist} & = & \frac{{\gamma\Delta{t}}}{1-c_2}\frac{\sqrt{c_2}}{1-\frac{\sqrt{c_2}}{4c_1}(\Omega_0\Delta{t})^2}\,. 
 \label{eq:SS_G_dist}
\end{eqnarray}\label{eq:SS_G}\noindent
\end{subequations}
It is obvious that none of the three linear benchmark characteristics in Eq.~(\ref{eq:SS_G}) show correct values for $\gamma\Delta{t}>0$, even if they all have the correct limiting values for $\gamma\Delta{t}\rightarrow0$. Figure~\ref{fig:fig_SS} shows the expressions of Eqs.~(\ref{eq:SS_G_diff}) and (\ref{eq:SS_G_drift}) in Fig.~\ref{fig:fig_SS}a, and Eq.~(\ref{eq:SS_G_dist}) in Fig.~\ref{fig:fig_SS}b, the latter for $\gamma=\frac{1}{2}\Omega_0$, $\gamma=\Omega_0$, $\gamma=\frac{3}{2}\Omega_0$, and $\gamma=\frac{5}{2}\Omega_0$.

Notice that the denominator of the last factor in Eq.~(\ref{eq:SS_G_dist}) reflects the stability criterion of Eq.~(\ref{eq:Stability_eq_R-}). Thus, the stability limit for any given friction coefficient, $\gamma$, is given by a singularity in $\Gamma_{\rm dist}$, as observed in Fig.~\ref{fig:fig_SS}b for $\gamma/\Omega_0=\frac{1}{2}$ and $\gamma=\Omega_0$. It follows from Eq.~(\ref{eq:SS_c2}) and either Eq.~(\ref{eq:SS_G_dist}) or (\ref{eq:Stability_eq_R-}) that a finite stability limit of the normalized time step, $\Omega_0\Delta{t}$, exists for this method if
\begin{eqnarray}
\frac{\gamma}{\Omega_0} & = & \frac{1}{Z}\ln\frac{1}{Z-\sqrt{Z^2-1}}  \label{eq:SS_stability_espression}
\end{eqnarray}
has a solution, in which case the resulting stability limit is given by
\begin{eqnarray}
\Omega_0\Delta{t} & < & 2\sqrt{Z} \,. \label{eq:SS_stability}
\end{eqnarray}
Equation~(\ref{eq:SS_stability_espression}) shows that $Z\ge1$, such that $\gamma\rightarrow0$ yields the stability limit $\Omega_0\Delta{t}<2$. The equation also shows that the method is stable for all time steps if $\gamma$ is above a certain threshold, exemplified in Fig.~\ref{fig:fig_SS}b for $\gamma/\Omega_0=\frac{3}{2}$ and $\gamma/\Omega_0=4$.

\subsection{{{EB$_{80}$}}}
\label{sec:sec_EB}

\begin{figure}[t]
\centering
\scalebox{0.585}{\centering \includegraphics[trim={3.10cm 13.75cm 1cm 5.0cm},clip]{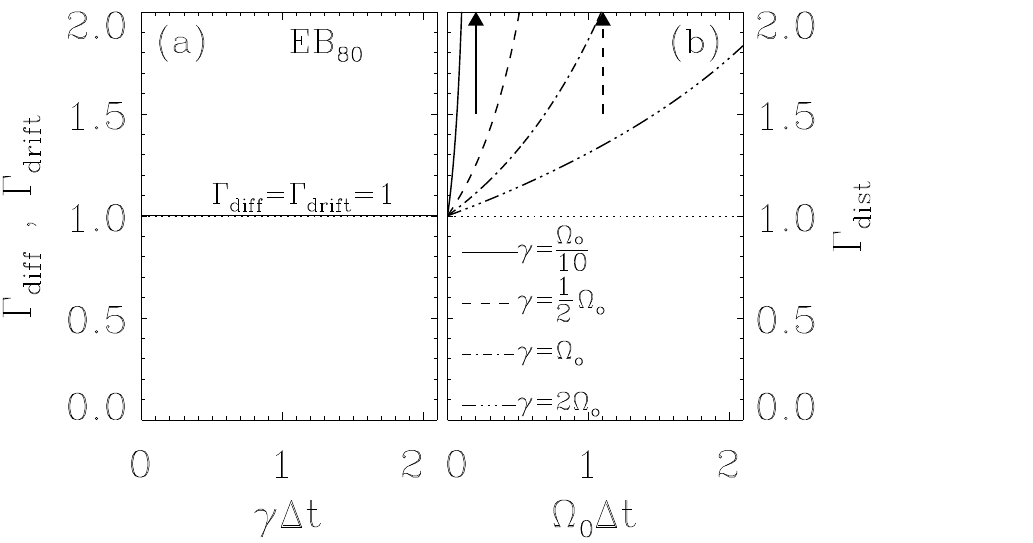}}
\caption{Normalized discrete-time diffusion [$\Gamma_{\rm diff}$, Eqs.~(\ref{eq:DE_discrete}) and (\ref{eq:EB_III_G_diff})], drift [$\Gamma_{\rm drift}$, Eqs.~(\ref{eq:Drift_discrete}) and (\ref{eq:EB_III_G_drift})], and configurational temperature [$\Gamma_{\rm dist}$, Eqs.~(\ref{eq:Dist_discrete}) and (\ref{eq:EB_III_G_dist})] for the EB$_{80}$ integrator of Ref.~\cite{Ermak1980}, as a function (a) of  $\gamma\Delta{t}$ for $\Gamma_{\rm diff}$ (solid) and $\Gamma_{\rm drift}$ (dashed), and (b) of $\Omega_0\Delta{t}$ for $\Gamma_{\rm dist}$, the latter for select values of normalized damping $\gamma/\Omega_0$ as indicated on the figure. Vertical arrows in (b) indicate the stability limit of the time step as given by Eqs.~(\ref{eq:Stability_eq_C}) and (\ref{eq:Stability_eq_R-}). Discrete-time quantities are normalized to the correct continuous-time quantities, Eq.~(\ref{eq:Basic_truths}).
}
\label{fig:fig_EB}
\end{figure}

Derived for constant force, the Ermak and Buckholtz EB$_{80}$ integrator \cite{Ermak1980} is originally given in three forms with $c_2=\exp(-\gamma\Delta{t})$, but it is most conveniently expressed in the form given in Eqs.~(27-34) of Ref.~\cite{Tanygin2024}:
\begin{subequations}
\begin{eqnarray}
r^{n+1} & = & r^n+\Delta{t}\frac{1-c_2}{\gamma\Delta{t}}\,v^n \nonumber \\
&& + \frac{\Delta{t}^2}{m}\frac{1}{\gamma\Delta{t}}\left(1-\frac{1-c_2}{\gamma\Delta{t}}\right)\,f^n + {\cal R}^n \\
v^{n+1} & = & c_2\,v^n + \frac{\Delta{t}}{m}\frac{1-c_2}{\gamma\Delta{t}}\,f^n + {\cal V}^n\,, 
\end{eqnarray}
where the stochastic variables, ${\cal R}^n$ and ${\cal V}^n$, are given by $\langle{\cal R}^n\rangle=\langle{\cal V}^n\rangle=0$, and
\begin{eqnarray}
\left\langle{\cal R}^n{\cal R}^\ell\right\rangle & = & \frac{k_BT}{m}\,\frac{2\Delta{t}}{\gamma}\left(1-\frac{3-c_2}{2}\frac{1-c_2}{\gamma\Delta{t}}\right)  \, \delta_{n,\ell}\label{eq:Ermak_rr}\\
\left\langle{\cal V}^n{\cal V}^\ell\right\rangle & = & \frac{k_BT}{m}\,(1-c_2^2) \, \delta_{n,\ell}\label{eq:Ermak_vv}\\
\left\langle{\cal R}^n{\cal V}^\ell\right\rangle & = & \frac{k_BT}{m}\,\gamma\Delta{t}^2\left(\frac{1-c_2}{\gamma\Delta{t}}\right)^2 \, \delta_{n,\ell}\,. \label{eq:Ermak_rv}
\end{eqnarray}\label{eq:Ermak}\noindent
\end{subequations}
Eliminating the velocity variable $v^n$ from Eq.~(\ref{eq:Ermak}) yields Eqs.~(\ref{eq:Vt_general}) and (\ref{eq:d_noise_all}) with the functional parameters:
\allowdisplaybreaks\begin{subequations}
\begin{eqnarray}
c_2 & = & 2c_1-1 \; = \; e^{-{\gamma\Delta{t}}} \label{eq:EB_III_c2}\\
c_3 & = & \frac{1}{\gamma\Delta{t}}\left(1-\frac{1-c_2}{{\gamma\Delta{t}}}\right)  \rightarrow \; \frac{1}{2} \; \; {\rm for} \; \; \gamma\Delta{t}\rightarrow0
\label{eq:EB_III_c3}  \\
c_4 & = & \frac{1}{\gamma\Delta{t}}\left({\frac{1-c_2}{{\gamma\Delta{t}}}-c_2}\right)  \rightarrow \; \frac{1}{2} \; \; {\rm for} \; \; \gamma\Delta{t}\rightarrow0 
\label{eq:EB_III_c4} \\
c_5 & = &  \frac{2}{{\gamma\Delta{t}}} \sqrt{c_2^2-(3c_1-2)\frac{1-c_2}{{\gamma\Delta{t}}}} 
 \rightarrow \frac{2}{\sqrt{3}} \;  {\rm for} \; \gamma\Delta{t}\rightarrow0\nonumber  \\ \label{eq:EB_III_c5}\\ 
c_6 & = &  \frac{2}{{\gamma\Delta{t}}} \sqrt{1-(2-c_1)\frac{1-c_2}{{\gamma\Delta{t}}}} 
  \rightarrow \; \frac{2}{\sqrt{3}} \; \; {\rm for} \; \; \gamma\Delta{t}\rightarrow0 \nonumber \\ \label{eq:EB_III_c6} \\
c_5c_6\zeta & = & \frac{4}{\left({\gamma\Delta{t}}\right)^2} \left(c_1\frac{1-c_2}{\gamma\Delta{t}}-c_2\right)  \rightarrow  \frac{1}{2} \; \; {\rm for} \; \gamma\Delta{t}\rightarrow0 \, . \nonumber\\ \label{eq:EB_III_zeta}
\end{eqnarray}\label{eq:EB_III_coeff}\noindent
\end{subequations}
Inserting these coefficients into Eqs.~(\ref{eq:DE_discrete}), (\ref{eq:Drift_discrete}), and (\ref{eq:Dist_discrete}) yields the results
\begin{subequations}
\begin{eqnarray}
\Gamma_{\rm diff} & = & 1 \label{eq:EB_III_G_diff}\\
\Gamma_{\rm drift} & = & 1 \label{eq:EB_III_G_drift}\\
\Gamma_{\rm dist} & = & \frac{1-c_2}{1-c_2-c_4(\Omega_0\Delta{t})^2}\displaystyle\frac{1+c_2\displaystyle\frac{c_3-c_4}{1-c_2^2}(\Omega_0\Delta{t})^2}{1-\frac{1}{2}\displaystyle\frac{c_3-c_4}{1+c_2}(\Omega_0\Delta{t})^2}  \,. \nonumber \\ \label{eq:EB_III_G_dist}
\end{eqnarray}\label{eq:EB_III_G}\noindent
\end{subequations}
The two transport characteristics, diffusion and drift, are correctly reproduced by the design of the method. The width of the sampling distribution, $\Gamma_{\rm dist}$ given in Eq.~(\ref{eq:EB_III_G_dist}), is strongly time-step dependent, limiting the usefulness of the method to very small time steps and appreciable damping parameters, even if the correct result is found for $\Delta{t}\rightarrow0$, as is expected.

Figure~\ref{fig:fig_EB} shows the expressions of Eqs.~(\ref{eq:EB_III_G_diff}) and (\ref{eq:EB_III_G_drift}) in Fig.~\ref{fig:fig_EB}a, and Eq.~(\ref{eq:EB_III_G_dist}) in Fig.~\ref{fig:fig_EB}b, the latter for $\gamma=\frac{1}{10}\Omega_0$, $\gamma=\frac{1}{2}\Omega_0$, $\gamma=\Omega_0$, and $\gamma=2\Omega_0$. The stability limits are indicated by the vertical arrows (two arrows are beyond the edge of the plot). As is obvious from Fig.~\ref{fig:fig_EB}b, this method has first order time step errors in $\Gamma_{\rm dist}$.

The method has the peculiar feature of having extremely limited stability for small $\gamma$. The reason is the stability criterion of Eq.~(\ref{eq:Stability_eq_C}), also reflected in the denominator of the first factor of Eq.~(\ref{eq:EB_III_G_dist}), which becomes problematic for $c_4$ of appreciable positive magnitudes. For this method $c_4\rightarrow\frac{1}{2}$ for $\gamma\Delta{t}\rightarrow0$, which yields an extremely small stability range for underdamped dynamics given that $c_2\rightarrow1-\gamma\Delta{t}$ in that same limit.

\subsection{{{MPA$_{80-82}$}}}
\label{sec:sec_REB}

\begin{figure}[t]
\centering
\scalebox{0.585}{\centering \includegraphics[trim={3.10cm 13.75cm 1cm 5.0cm},clip]{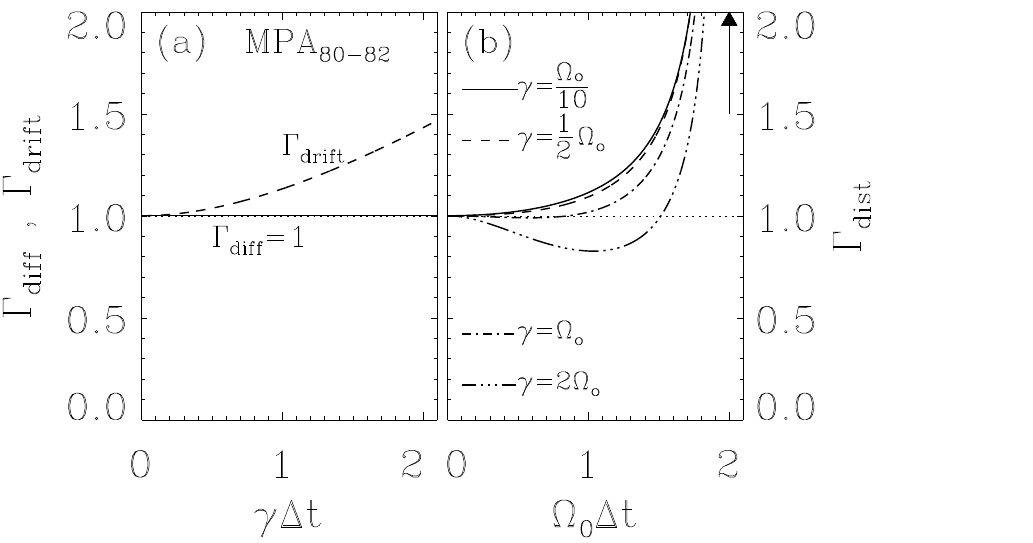}}
\caption{Normalized discrete-time diffusion [$\Gamma_{\rm diff}$, Eqs.~(\ref{eq:DE_discrete}) and (\ref{eq:REB_III_G_diff})], drift [$\Gamma_{\rm drift}$, Eqs.~(\ref{eq:Drift_discrete}) and (\ref{eq:REB_III_G_drift})], and Boltzmann configurational temperature [$\Gamma_{\rm dist}$, Eqs.~(\ref{eq:Dist_discrete}) and (\ref{eq:REB_III_G_dist})] for the MPA$_{80-82}$ integrator of Ref.~\cite{Allen80,Allen82}, as a function (a) of  $\gamma\Delta{t}$ for $\Gamma_{\rm diff}$ (solid) and $\Gamma_{\rm drift}$ (dashed), and (b) of $\Omega_0\Delta{t}$ for $\Gamma_{\rm dist}$, the latter for select values of normalized damping $\gamma/\Omega_0$ as indicated on the figure. Vertical arrow in (b) indicates the stability limit, $\Omega_0\Delta{t}<2$, of the time step as given by Eq.~(\ref{eq:Stability_eq_R-}). Discrete-time quantities are normalized to the correct continuous-time quantities, Eq.~(\ref{eq:Basic_truths}).
}
\label{fig:fig_REB}
\end{figure}

Addressing the constant force assumption in EB$_{80}$, the MPA$_{80-82}$ integrator is given in Eqs.~(2)-(4) of Ref.~\cite{Thalmann2007} and attributed to Allen \cite{Allen80,Allen82}:
\begin{subequations}
\begin{eqnarray}
r^{n+1} & = & r^n+\Delta{t}\frac{1-c_2}{\gamma\Delta{t}}\,v^n + \frac{\Delta{t}^2}{2m}\,f^n + {\cal R}^n \\
v^{n+1} & = & c_2\,v^n + \frac{\Delta{t}}{2m}\frac{1-c_2}{\gamma\Delta{t}}\,(f^n+f^{n+1}) + {\cal V}^n\,,
\end{eqnarray}\label{eq:Allen}\noindent
\end{subequations}
where the stochastic variables, ${\cal R}^n$ and ${\cal V}^n$, are given by $\langle{\cal R}^n\rangle=\langle{\cal V}^n\rangle=0$ and Eqs.~(\ref{eq:Ermak_rr}), (\ref{eq:Ermak_vv}), and (\ref{eq:Ermak_rv}).
Eliminating the velocity variable $v^n$ from Eq.~(\ref{eq:Allen}) yields Eq.~(\ref{eq:Vt_general}) with the functional parameters:
\begin{subequations}
\begin{eqnarray}
c_2 &= & 2c_1-1 \; = \; e^{-{\gamma\Delta{t}}} \label{eq:REB_III_c2}\\
c_3 & = & \frac{1}{2}\left[\left(\frac{1-c_2}{\gamma\Delta{t}}\right)^2+1\right]  \; \rightarrow \; 1 \; \; {\rm for} \; \; \gamma\Delta{t}\rightarrow0 \label{eq:REB_III_c3} \\
c_4 & = & \frac{1}{2}\left[\left(\frac{1-c_2}{\gamma\Delta{t}}\right)^2-c_2\right]  \; \rightarrow \; 0 \; \; {\rm for} \; \; \gamma\Delta{t}\rightarrow0\label{eq:REB_III_c4} \\
c_5 & = &  \frac{2}{{\gamma\Delta{t}}} \sqrt{c_2^2-(3c_1-2)\frac{1-c_2}{{\gamma\Delta{t}}}}  \; \rightarrow \; \frac{2}{\sqrt{3}} \; \; {\rm for} \; \; \gamma\Delta{t}\rightarrow0 \nonumber \\ \label{eq:REB_c5} \\
c_6 & = &  \frac{2}{{\gamma\Delta{t}}} \sqrt{1-(2-c_1)\frac{1-c_2}{{\gamma\Delta{t}}}}   \; \rightarrow \; \frac{2}{\sqrt{3}} \; \; {\rm for} \; \; \gamma\Delta{t}\rightarrow0 \nonumber \\
\label{eq:REB_III_c6} \\
c_5c_6\zeta & = & \frac{4}{(\gamma\Delta{t})^2} \left(c_1\frac{1-c_2}{{\gamma\Delta{t}}}-c_2\right)  \; \rightarrow \; \frac{1}{2} \; \; {\rm for} \; \; \gamma\Delta{t}\rightarrow0\,, \nonumber \\
\label{eq:REB_zeta}
\end{eqnarray}\label{eq:RE_coeff}\noindent
\end{subequations}
where the noise parameters, $c_5$, $c_6$, and $\zeta$, are the same as Eqs.~(\ref{eq:EB_III_c5}), (\ref{eq:EB_III_c6}), and (\ref{eq:EB_III_zeta}). Inserting these coefficients into Eqs.~(\ref{eq:DE_discrete}), (\ref{eq:Drift_discrete}), and (\ref{eq:Dist_discrete}) yields the results
\begin{subequations}
\begin{eqnarray}
&&\Gamma_{\rm diff} \; = \; 1 \label{eq:REB_III_G_diff}\\
&&\Gamma_{\rm drift} \; = \; \frac{1-c_2}{\gamma\Delta{t}}+\frac{\gamma\Delta{t}}{2} \label{eq:REB_III_G_drift}\\
&&\Gamma_{\rm dist} \; = \; \frac{\gamma\Delta{t}}{1-c_2-c_4(\Omega_0\Delta{t})^2} \; \times \label{eq:REB_III_G_dist}\\
 &&\displaystyle\frac{8c_1\left(\frac{1-c_2}{\gamma\Delta{t}}\right)^2+\left[4c_4\left(\frac{1-c_2}{\gamma\Delta{t}}\right)^2-2(c_3+c_4)c_5c_6\zeta\right](\Omega_0\Delta{t})^2}{8c_1(c_3+c_4)[1-\frac{1}{4}(\Omega_0\Delta{t})^2]} . \nonumber
\end{eqnarray}\label{eq:REB_III_G}\noindent
\end{subequations}
Figure~\ref{fig:fig_REB} shows the expressions of Eqs.~(\ref{eq:REB_III_G_diff}) and (\ref{eq:REB_III_G_drift}) in Fig.~\ref{fig:fig_REB}a, and Eq.~(\ref{eq:REB_III_G_dist}) in Fig.~\ref{fig:fig_REB}b, the latter for $\gamma=\frac{1}{10}\Omega_0$, $\gamma=\frac{1}{2}\Omega_0$, $\gamma=\Omega_0$, and $\gamma=2\Omega_0$. The stability limit, $\Omega_0\Delta{t}<2$, from Eq.~(\ref{eq:Stability_eq_R-}) is indicated by the vertical arrow in Fig.~\ref{fig:fig_REB}b. As seen from the figure, despite the normalized drift velocity $\Gamma_{\rm drift}>0$, this method is a significant improvement over EB$_{80}$ in Sec.~\ref{sec:sec_EB} due to the much better configurational sampling demonstrated in Eq.~(\ref{eq:REB_III_G_dist}), and the unproblematic stability behavior for $\gamma\rightarrow0$ compared to the observed behavior for EB$_{80}$ in the same limit.

\subsection{{{vGB$_{82}$}}}
\label{sec:sec_vGB82}

\begin{figure}[t]
\centering
\scalebox{0.585}{\centering \includegraphics[trim={3.10cm 13.75cm 1cm 5.0cm},clip]{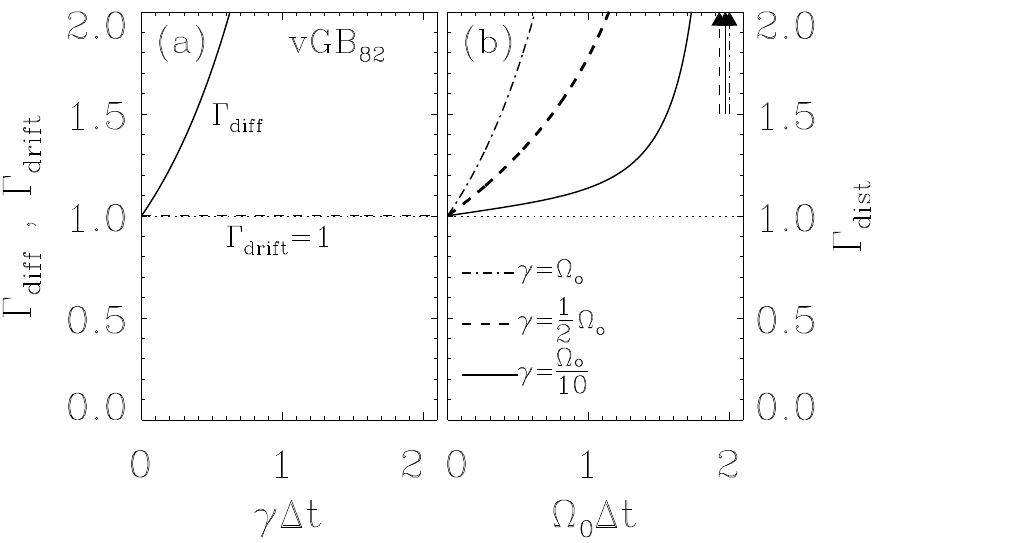}}
\caption{Normalized discrete-time diffusion [$\Gamma_{\rm diff}$, Eqs.~(\ref{eq:DE_discrete}) and (\ref{eq:vg82_G_diff})], drift [$\Gamma_{\rm drift}$, Eqs.~(\ref{eq:Drift_discrete}) and (\ref{eq:vg82_G_drift})], and Boltzmann configurational temperature [$\Gamma_{\rm dist}$, Eqs.~(\ref{eq:Dist_discrete}) and (\ref{eq:vg82_G_dist})] for the vGB$_{82}$ integrator of Ref.~\cite{vGB82}, as a function (a) of  $\gamma\Delta{t}$ for $\Gamma_{\rm diff}$ (solid) and $\Gamma_{\rm drift}$ (dashed), and (b) of $\Omega_0\Delta{t}$ for $\Gamma_{\rm dist}$, the latter for select values of normalized damping $\gamma/\Omega_0$ as indicated on the figure. Vertical arrows in (b) indicate the stability limit of the time step as given by Eq.~(\ref{eq:Stability_eq_R-}). Discrete-time quantities are normalized to the correct continuous-time quantities, Eq.~(\ref{eq:Basic_truths}).
}
\label{fig:fig_vGB82}
\end{figure}

Similarly to the MPA$_{80-82}$ integrator of Allen, van Gunsteren and Berendsen addressed the constant force approximation of the EB$_{80}$ method with the vGB$_{82}$ integrator \cite{vGB82}. The algorithm originates in the configurational form of Eq.~(\ref{eq:Vt_general}) and the functional coefficients are:
\begin{subequations}
\begin{eqnarray}
c_2 & = & 2c_1-1\; = \; e^{-{\gamma\Delta{t}}} \label{eq:vg82_c2}\\
c_3 & = &  \frac{1-c_2}{{\gamma\Delta{t}}}+\frac{1}{\gamma\Delta{t}}\left[c_1-\frac{1-c_2}{{\gamma\Delta{t}}}\right]   \; \rightarrow \; 1 \; \; {\rm for} \; \; \gamma\Delta{t}\rightarrow0 \nonumber \\
\label{eq:vg82_c3} \\
c_4 & = & -\frac{1}{\gamma\Delta{t}}\left[c_1-\frac{1-c_2}{{\gamma\Delta{t}}}\right]   \; \rightarrow \; 0 \; \; {\rm for} \; \; \gamma\Delta{t}\rightarrow0\label{eq:vg82_c4} \\
c_5 & = &  \frac{1}{c_2}\frac{2}{\gamma\Delta{t}}\sqrt{c_2^2-(2c_2-c_1)\frac{1-c_2}{{\gamma\Delta{t}}}}  \label{eq:vg82_c5}\\
&& \; \rightarrow \; \frac{2}{\sqrt{3}} \; \; {\rm for} \; \; \gamma\Delta{t}\rightarrow0\nonumber \\
c_6 & = &  \frac{2}{\gamma\Delta{t}}\sqrt{1-(2-c_1)\frac{1-c_2}{{\gamma\Delta{t}}}}  \; \rightarrow \; \frac{2}{\sqrt{3}} \; \; {\rm for} \; \; \gamma\Delta{t}\rightarrow0\nonumber \\
\label{eq:vg82_c6} \\
c_5c_6\zeta & = &  \frac{1}{c_2}\frac{4}{(\gamma\Delta{t})^{2}}\left(c_1\frac{1-c_2}{{\gamma\Delta{t}}}-c_2\right)   \; \rightarrow \; \frac{1}{2} \; \; {\rm for} \; \; \gamma\Delta{t}\rightarrow0.\nonumber \\
\label{eq:vg82_zeta}
\end{eqnarray}\label{eq:vg82_coeff}\noindent
\end{subequations}
Inserting these coefficients into Eqs.~(\ref{eq:DE_discrete}), (\ref{eq:Drift_discrete}), and (\ref{eq:Dist_discrete}) yields the normalized quantities
\begin{subequations}
\begin{eqnarray}
&&\Gamma_{\rm diff} \; = \;  \frac{c_1}{c_2^2}\frac{1-c_2}{{\gamma\Delta{t}}} \label{eq:vg82_G_diff}\\
&&\Gamma_{\rm drift} \; = \; 1 \label{eq:vg82_G_drift}\\
&&\Gamma_{\rm dist} \; = \; \frac{1-c_2}{1-c_2-c_4(\Omega_0\Delta{t})^2} \left(\frac{1-c_2}{\gamma\Delta{t}}\right) \; \times \label{eq:vg82_G_dist}\\
&&\frac{4c_1\left[\frac{c_1}{c_2^2}\right]+\left[2\frac{c_4c_1}{c_2^2}-\frac{c_5c_6\zeta}{(\frac{1-c_2}{\gamma\Delta{t}})^2}\right](\Omega_0\Delta{t})^2}{4c_1-(c_3-c_4)\,(\Omega_0\Delta{t})^2} \,. \nonumber 
\end{eqnarray}\label{eq:vGB82_G}\noindent
\end{subequations}
Figure~\ref{fig:fig_vGB82} shows the expressions of Eqs.~(\ref{eq:vg82_G_diff}) and (\ref{eq:vg82_G_drift}) in Fig.~\ref{fig:fig_vGB82}a, and Eq.~(\ref{eq:vg82_G_dist}) in Fig.~\ref{fig:fig_vGB82}b, the latter for $\gamma=\frac{1}{10}\Omega_0$, $\gamma=\frac{1}{2}\Omega_0$, and $\gamma=\Omega_0$. The stability limit from Eq.~(\ref{eq:Stability_eq_R-}) is indicated by the vertical arrows in Fig.~\ref{fig:fig_vGB82}b. As seen from the figure and Eq.~(\ref{eq:vGB82_G}), the normalized drift velocity, $\Gamma_{\rm drift}=1$, is precise while the normalized diffusion, $\Gamma_{\rm diff}>1$, is rapidly deviating from the correct, continuous time result for increasing time step. The behavior of the normalized width, $\Gamma_{\rm dist}$, of the sampling distribution is not directly transparent from either Eq.~(\ref{eq:vg82_G_dist}) or (\ref{eq:Dist_discrete}), but Fig.~\ref{fig:fig_vGB82}b indicates an improvement over EB$_{80}$ in that low damping values result in a reasonable range for the normalized time step, $\Omega_0\Delta{t}$, within which the sampling error is somewhat confined, even if the error seems to be of first order.

\subsection{{{BBK$_{84}$}}}
\label{sec:sec_BBK}

\begin{figure}[t]
\centering
\scalebox{0.585}{\centering \includegraphics[trim={3.10cm 13.75cm 1cm 5.0cm},clip]{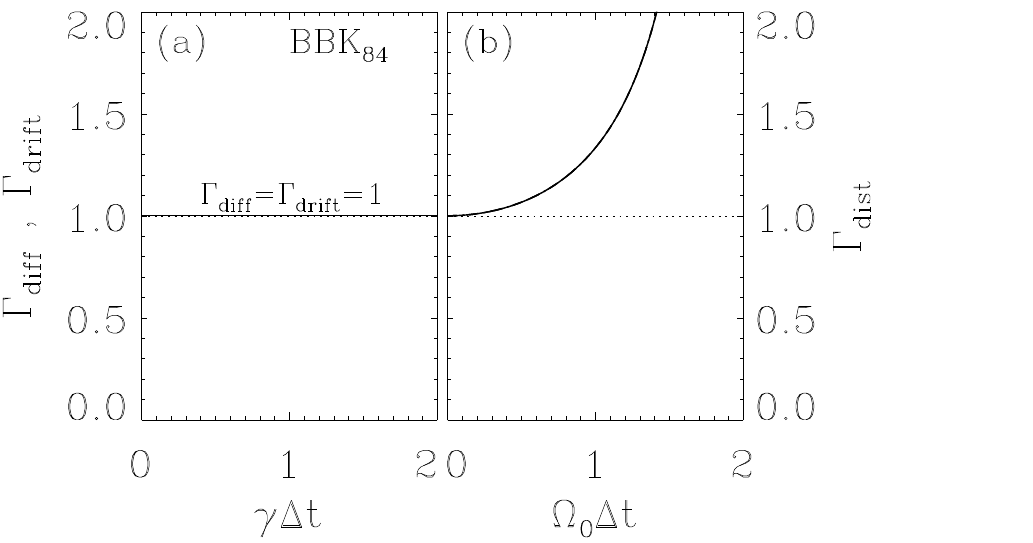}}
\caption{Normalized discrete-time diffusion [$\Gamma_{\rm diff}$, Eqs.~(\ref{eq:DE_discrete}) and (\ref{eq:BBK_G_diff})], drift [$\Gamma_{\rm drift}$, Eqs.~(\ref{eq:Drift_discrete}) and (\ref{eq:BBK_G_drift})], and Boltzmann configurational temperature [$\Gamma_{\rm dist}$, Eqs.~(\ref{eq:Dist_discrete}) and (\ref{eq:BBK_G_dist})] for the BBK$_{84}$ integrator of Ref.~\cite{BBK}, as a function (a) of  $\gamma\Delta{t}$ for $\Gamma_{\rm diff}$ (solid) and $\Gamma_{\rm drift}$ (dashed), and (b) of $\Omega_0\Delta{t}$ for $\Gamma_{\rm dist}$, the latter for any value of normalized damping $\gamma/\Omega_0>0$. The stability limit of the time step is given by Eq.~(\ref{eq:Stability_eq_R-}) to be $\Omega_0\Delta{t}<2$. Discrete-time quantities are normalized to the correct continuous-time quantities, Eq.~(\ref{eq:Basic_truths}).
}
\label{fig:fig_BBK}
\end{figure}

The BBK$_{84}$ integrator \cite{BBK} by Br{\"u}nger, Brooks, and Karplus, also analyzed in Ref.~\cite{Pastor_88}, is widely used through the ``fix~langevin" option in the molecular modeling suite LAMMPS \cite{Plimpton2}, and it is used in the ESPResSo suite \cite{Espresso_4.2.0}, both implementations seemingly applying uniformly distributed noise variables. The integrator originates in the configurational form of Eq.~(\ref{eq:Vt_general}) where the functional coefficients can be written:
\begin{subequations}
\begin{eqnarray}
c_2 & = & 2c_1-1 \; = \; \frac{1-{\frac{1}{2}\gamma\Delta{t}}}{1+{\frac{1}{2}\gamma\Delta{t}}} \label{eq:BBK_c2}\\
c_3 & = & c_1 \; = \; \frac{1}{1+{\frac{1}{2}\gamma\Delta{t}}}   \; \rightarrow \; 1 \; \; {\rm for} \; \; \gamma\Delta{t}\rightarrow0\label{eq:BBK_c3} \\
c_4 & = & 0 \label{eq:BBK_c4} \\
c_5 & = & 0 \label{eq:BBK_c5} \\
c_6 & = & 2c_1 \; \rightarrow \; 2 \; \; {\rm for} \; \; \gamma\Delta{t}\rightarrow0 \label{eq:BBK_c6}   \\
\zeta & = & 0\, . \label{eq:BBK_zeta}
\end{eqnarray}\label{eq:BBK_coeff}\noindent
\end{subequations}
Inserting these coefficients into Eqs.~(\ref{eq:DE_discrete}), (\ref{eq:Drift_discrete}), and Eq.~(\ref{eq:Dist_discrete}), yields the results
\begin{subequations}
\begin{eqnarray}
\Gamma_{\rm diff} & = & 1 \label{eq:BBK_G_diff}\\
\Gamma_{\rm drift} & = & 1 \label{eq:BBK_G_drift}\\
\Gamma_{\rm dist} & = & \frac{1}{1-\left(\frac{\Omega_0\Delta{t}}{2}\right)^2} \, . \label{eq:BBK_G_dist}
\end{eqnarray}\label{eq:BBK_G}\noindent
\end{subequations}
In this case, the two transport characteristics, diffusion and drift, of the three linear benchmark characteristics in Eq.~(\ref{eq:BBK_G}), show correct values for all time-steps $\Omega_0\Delta{t}<2$, as given by the stability criterion of Eq.~(\ref{eq:Stability_eq_R-}). Figure~\ref{fig:fig_BBK} shows the expressions of Eqs.~(\ref{eq:BBK_G_diff}) and (\ref{eq:BBK_G_drift}) in Fig.~\ref{fig:fig_BBK}a and Eq.~(\ref{eq:BBK_G_dist}) in Fig.~\ref{fig:fig_BBK}b, the latter for any value of $\gamma>0$. The width of the sampling distribution is represented by $\Gamma_{\rm dist}$ in Eq.~(\ref{eq:BBK_G_dist}), and is always larger than the correct value, diverging when $\Omega_0\Delta{t}$ approaches the stability limit. This is consistent with the BBK$_{84}$ method exhibiting elevated configurational temperature for increasing time steps.

\subsection{{{vGB$_{88}$}}}
\label{sec:sec_vGB88}

\begin{figure}[t]
\centering
\scalebox{0.585}{\centering \includegraphics[trim={3.10cm 13.75cm 1cm 5.0cm},clip]{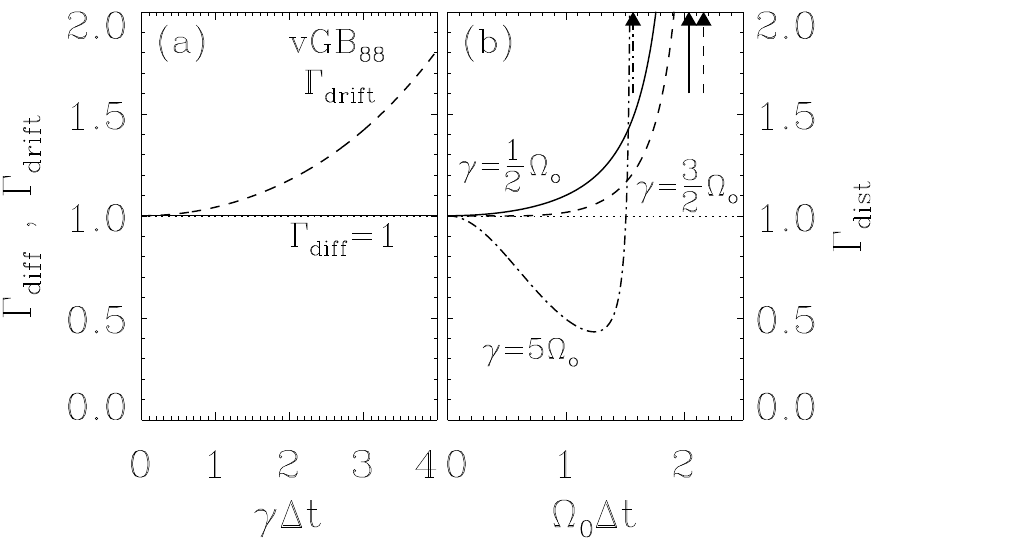}}
\caption{Normalized discrete-time diffusion [$\Gamma_{\rm diff}$, Eqs.~(\ref{eq:DE_discrete}) and (\ref{eq:vGB88_G_diff})], drift [$\Gamma_{\rm drift}$, Eqs.~(\ref{eq:Drift_discrete}) and (\ref{eq:vGB88_G_drift})], and Boltzmann configurational temperature [$\Gamma_{\rm dist}$, Eqs.~(\ref{eq:Dist_discrete}) and (\ref{eq:vGB88_G_dist})] for the vGB$_{88}$ integrator of Ref.~\cite{vGB88}, as a function (a) of  $\gamma\Delta{t}$ for $\Gamma_{\rm diff}$ (solid) and $\Gamma_{\rm drift}$ (dashed), and (b) of $\Omega_0\Delta{t}$ for $\Gamma_{\rm dist}$, the latter for select values of normalized damping $\gamma/\Omega_0$ as indicated on the figure. Vertical arrows in (b) indicate the stability limit of the time step as given by Eq.~(\ref{eq:Stability_eq_R-}). Discrete-time quantities are normalized to the correct continuous-time quantities, Eq.~(\ref{eq:Basic_truths}).
}
\label{fig:fig_vGB88}
\end{figure}

With $c_2=\exp(-\gamma\Delta{t})$, the vGB$_{88}$ integrator \cite{vGB88} by van Gunsteren and Berendsen is a leap-frog formulation of a Verlet-type method with a very careful consideration of the noise, which is  resolved for each half time step. The method reads in its original form:
\begin{subequations}
\begin{eqnarray}
v^{n+\frac{1}{2}} & = & c_2\,v^{n-\frac{1}{2}} + \frac{\Delta{t}}{m}\frac{1-c_2}{\gamma\Delta{t}}f^n +{\cal V}_+^n-c_2{\cal V}_-^n\\
r^{n+1} & = & r^n + \Delta{t}\frac{1}{\sqrt{c_2}}\frac{1-c_2}{\gamma\Delta{t}}\,v^{n+\frac{1}{2}} + {\cal R}_+^{n+\frac{1}{2}}-{\cal R}_-^{n+\frac{1}{2}} \, , \nonumber \\
\end{eqnarray}
where the stochastic variables, ${\cal R}_\pm^{n+\frac{1}{2}}$ and ${\cal V}_\pm^n$, are given by
\begin{eqnarray}
{\cal V}_-^{n} & = & -\frac{1}{m\sqrt{c_2}}\int_{t_n-\frac{\Delta{t}}{2}}^{t_n}e^{\gamma(t-t_n)}\beta(t)\,dt \\
{\cal V}_+^{n} & = & \frac{\sqrt{c_2}}{m}\int_{t_n}^{t_n+\frac{\Delta{t}}{2}}e^{\gamma(t-t_n)}\beta(t)\,dt \\
{\cal R}_-^{n+\frac{1}{2}} & = & -\frac{1}{m\gamma}\int_{t_n}^{t_n+\frac{\Delta{t}}{2}}\left[1-e^{\gamma(t-t_n)}\right]\beta(t)\,dt \\
{\cal R}_+^{n+\frac{1}{2}} & = & \frac{1}{m\gamma}\int_{t_n+\frac{\Delta{t}}{2}}^{t_{n+1}}\left[1-e^{\gamma(t-t_{n+1})}\right]\beta(t)\,dt \, .
\end{eqnarray}\label{eq:vGB88_orig}\noindent
\end{subequations}
Eliminating the velocity variable $v^{n+\frac{1}{2}}$ from Eq.~(\ref{eq:vGB88_orig}) yields Eq.~(\ref{eq:Vt_general}) with the functional parameters:
\begin{subequations}
\begin{eqnarray}
c_2 & = & 2c_1-1 \; = \; e^{-{\gamma\Delta{t}}} \label{eq:vGB88_c2}\\
c_3 & = & \frac{1}{\sqrt{c_2}}\left(\frac{1-c_2}{{\gamma\Delta{t}}}\right)^2\; \rightarrow \; 1 \; \; {\rm for} \; \; \gamma\Delta{t}\rightarrow0 \label{eq:vGB88_c3} \\
c_4 & = & 0 \label{eq:vGB88_c4} \\
c_5 & = & \frac{2}{{\gamma\Delta{t}}}\sqrt{c_2^2-(2c_2-c_1)\frac{1-c_2}{{\gamma\Delta{t}}}} \; \rightarrow \; \frac{2}{\sqrt{3}} \; \; {\rm for} \; \; \gamma\Delta{t}\rightarrow0\nonumber \\
\label{eq:vGB88_c5} \\
c_6 & = &  \frac{2}{{\gamma\Delta{t}}}\sqrt{1-(2-c_1)\frac{1-c_2}{{\gamma\Delta{t}}}}\; \rightarrow \; \frac{2}{\sqrt{3}} \; \; {\rm for} \; \; \gamma\Delta{t}\rightarrow0 \nonumber \\
\label{eq:vGB88_c6} \\
c_5c_6\zeta & = &   \frac{4}{\left({\gamma\Delta{t}}\right)^2} \left(c_1\frac{1-c_2}{{\gamma\Delta{t}}}-c_2\right)\; \rightarrow \; \frac{1}{2} \; \; {\rm for} \; \; \gamma\Delta{t}\rightarrow0 \, , \nonumber \\
\label{eq:vGB88_zeta} 
\end{eqnarray} \label{eq:vGB88_coeff}\noindent
\end{subequations}
where we see that the four stochastic variables from Eq.~(\ref{eq:vGB88_orig}) are combined into the same two variables used in the EB$_{80}$ and MPA$_{80}$ integrators. Inserting these coefficients into Eqs.~(\ref{eq:DE_discrete}), (\ref{eq:Drift_discrete}), and Eq.~(\ref{eq:Dist_discrete}), yields the results
\begin{subequations}
\begin{eqnarray}
\Gamma_{\rm diff} & = & 1 \label{eq:vGB88_G_diff}\\
\Gamma_{\rm drift} & = &  \frac{1}{\sqrt{c_2}}\frac{1-c_2}{{\gamma\Delta{t}}} \label{eq:vGB88_G_drift}\\
\Gamma_{\rm dist} & = & \frac{{\gamma\Delta{t}}}{1-c_2}\sqrt{c_2}\frac{1-\frac{c_5c_6\zeta}{4c_1\sqrt{c_2}}(\Omega_0\Delta{t})^2}{1-\frac{c_3}{4c_1}(\Omega_0\Delta{t})^2} \,.\label{eq:vGB88_G_dist}
\end{eqnarray}\label{eq:vGB88_G}\noindent
\end{subequations}
Figure~\ref{fig:fig_vGB88} shows the expressions of Eqs.~(\ref{eq:vGB88_G_diff}) and (\ref{eq:vGB88_G_drift}) in Fig.~\ref{fig:fig_vGB88}a, and Eq.~(\ref{eq:vGB88_G_dist}) in Fig.~\ref{fig:fig_vGB88}b, the latter for $\gamma=\frac{1}{2}\Omega_0$, $\gamma=\frac{3}{2}\Omega_0$, and $\gamma=5\Omega_0$. The stability limit from Eq.~(\ref{eq:Stability_eq_R-}) is indicated by the vertical arrows in Fig.~\ref{fig:fig_vGB88}b. As seen from the figure and Eq.~(\ref{eq:vGB88_G}), the normalized drift velocity, $\Gamma_{\rm drift}$, is increasingly elevated, while the normalized diffusion is correct for all time steps. The behavior of the normalized width, $\Gamma_{\rm dist}$, of the sampling distribution is given from either Eq.~(\ref{eq:vGB88_G_dist}) or (\ref{eq:Dist_discrete}), and Fig.~\ref{fig:fig_vGB88}b indicates an improvement over vGB$_{82}$ in that the error seems to be of second order in the time step.

\subsection{{{GW$_{97}$}}}
\label{sec:sec_GW97}

\begin{figure}[t]
\centering
\scalebox{0.585}{\centering \includegraphics[trim={3.10cm 13.75cm 1cm 5.0cm},clip]{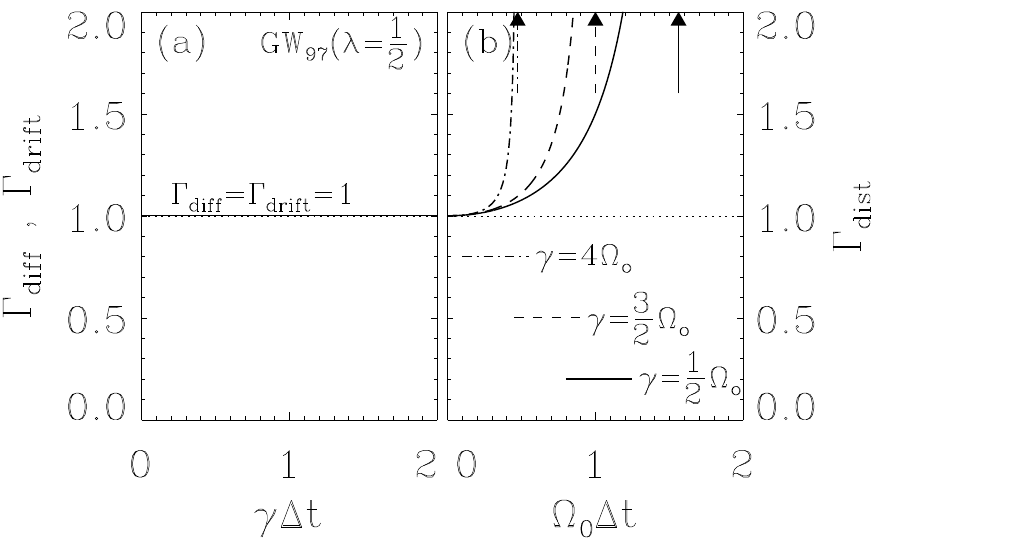}}
\caption{Normalized discrete-time diffusion [$\Gamma_{\rm diff}$, Eqs.~(\ref{eq:DE_discrete}) and (\ref{eq:GW97_G_diff})], drift [$\Gamma_{\rm drift}$, Eqs.~(\ref{eq:Drift_discrete}) and (\ref{eq:GW97_G_drift})], and Boltzmann configurational temperature [$\Gamma_{\rm dist}$, Eqs.~(\ref{eq:Dist_discrete}) and (\ref{eq:GW97_G_dist})] for the GW$_{97}$ integrator of Ref.~\cite{GW97} with $\lambda=\frac{1}{2}$, as a function (a) of  $\gamma\Delta{t}$ for $\Gamma_{\rm diff}$ (solid) and $\Gamma_{\rm drift}$ (dashed), and (b) of $\Omega_0\Delta{t}$ for $\Gamma_{\rm dist}$, the latter for select values of normalized damping $\gamma/\Omega_0$ as indicated on the figure. Vertical arrows in (b) indicate the stability limit of the time step as given by Eq.~(\ref{eq:Stability_GW}). Discrete-time quantities are normalized to the correct continuous-time quantities, Eq.~(\ref{eq:Basic_truths}).
}
\label{fig:fig_GW97}
\end{figure}

The GW$_{97}$ integrator by Groot and Warren \cite{GW97} reads:
\begin{subequations}
\begin{eqnarray}
r^{n+1} & = & r^n+\Delta{t}\,v^n+\frac{\Delta{t}^2}{2m}f^n-\frac{\Delta{t}^2}{2m}\alpha\,\tilde{v}^{n-1+\lambda}+\frac{\Delta{t}}{2m}\beta_+^{n} \label{eq:GW97_1} \nonumber \\ \\
\tilde{v}^{n+\lambda} & = & v^n+\frac{\lambda}{m} [\Delta{t} (f^n-\alpha\,\tilde{v}^{n-1+\lambda})+\beta_+^n] \label{eq:GW97_2} \\
v^{n+1} & = & v^n+\frac{\Delta{t}}{2m}(f^n+f^{n+1}) \label{eq:GW97_3} \\
&& -\frac{\Delta{t}}{2m}\alpha(\tilde{v}^{n-1+\lambda}+\tilde{v}^{n+\lambda}) + \frac{1}{2m}(\beta_+^n+\beta_+^{n+1})\, ,\nonumber 
\end{eqnarray}\label{eq:GW97_1-3}\noindent
\end{subequations}
where $\lambda$ is a free parameter that in Ref.~\cite{GW97} is suggested to be $\lambda=\frac{1}{2}$. Notice that $\tilde{v}^{n+\lambda}$ is denoted $\tilde{v}^{n+1}$ in Ref.~\cite{GW97}, indicating that this velocity is an approximation at time $t_{n+1}$ for any $\lambda$. However, as intuitively apparent from Eq.~(\ref{eq:GW97_2}), this is not the case in general, and certainly not for $\lambda=0$. In more detail, combining Eq.~(\ref{eq:GW97_1}) with Eq.~(\ref{eq:GW97_2}) yields
\begin{eqnarray}
\tilde{v}^{n+\lambda} & = & (1-2\lambda)v^n+2\lambda\frac{r^{n+1}-r^n}{\Delta{t}} \label{eq:GW97_tilde_v}
\end{eqnarray}
with
\begin{eqnarray}
v^n & = & \frac{r^{n+1}-r^{n-1}}{2\Delta{t}} \,, \label{eq:GW97_vn}
\end{eqnarray}
the latter obtained by combining Eqs.~(\ref{eq:GW97_1}) and (\ref{eq:GW97_3}).
Thus, we see from the statistical requirement of anti-symmetry with $r^n$ for a half-step velocity \cite{2GJ,GJ,GJ24} that
\begin{eqnarray}
\langle(r^n+r^{n+1})\tilde{v}^{n+\lambda}\rangle  =  \left\langle(\tilde{v}^{n-\frac{1}{2}}+\tilde{v}^{n+\frac{1}{2}})r^n\right\rangle & = & 0 \,, \, {\rm for} \; \lambda=\frac{1}{2} \nonumber \\\label{eq:GW97_HS}
\end{eqnarray}
and, from the statistical requirement for a velocity on-site with $r^n$ \cite{GJ,GJ24}, that
\begin{eqnarray}
\langle r^n\tilde{v}^{n+\lambda}\rangle & = & 0 \; \; , \; \; {\rm for} \; \lambda=0 \, .\label{eq:GW97_OS}
\end{eqnarray}
It is therefore apparent that the velocity $\tilde{v}^{n+\lambda}$ is a half-step velocity for $\lambda=\frac{1}{2}$, and for this value of $\lambda$ the friction force in Eq.~(\ref{eq:GW97_1-3}) is lagging a half time-step compared to the expectation by the original notation used for the integrator. We notice parenthetically that, while $\tilde{v}^{n+\lambda}$ is on-site with $r^n$ for $\lambda=0$, $\lambda=1$ does not lead to a velocity, $\tilde{v}^{n+1}$, on-site with $r^{n+1}$.

Writing the configurational equation from Eqs.~(\ref{eq:GW97_1}), (\ref{eq:GW97_tilde_v}), and (\ref{eq:GW97_vn}) we obtain
\begin{subequations}
\begin{eqnarray}
r^{n+1} & = & 2(1-\frac{1+2\lambda}{2}{\frac{1}{2}\gamma\Delta{t}})r^n-(1-4\lambda{\frac{1}{2}\gamma\Delta{t}})r^{n-1}\nonumber \\
&&+{\frac{1}{2}\gamma\Delta{t}}(1-2\lambda)r^{n-2} +\frac{\Delta{t}^2}{m}f^n+\frac{\Delta{t}}{m}\beta_+^n \,,\label{eq:GW97_rr}
\end{eqnarray}
and the corresponding velocity equation reads from Eqs.~(\ref{eq:GW97_3}), (\ref{eq:GW97_tilde_v}), and (\ref{eq:GW97_vn}):
\begin{eqnarray}
v^{n+1} & = & (1-(1+2\lambda){\frac{1}{2}\gamma\Delta{t}})v^n-(1-2\lambda){\frac{1}{2}\gamma\Delta{t}}v^{n-1}\nonumber \\
&&+\frac{\Delta{t}}{2m}(f^n+f^{n+1}) + \frac{1}{2m}(\beta_+^n+\beta_+^{n+1})\, . \label{eq:GW97_vv}
\end{eqnarray}\label{eq:GW97_rrvv}\noindent
\end{subequations}
Clearly, for $\lambda\neq\frac{1}{2}$, Eq.~(\ref{eq:GW97_rr}) does not conform to Eq.~(\ref{eq:Vt_general}) due to the term $\propto r^{n-2}$. Indeed, the properties of the set of Eq.~(\ref{eq:GW97_rrvv}) are peculiar for $\lambda\neq\frac{1}{2}$. One example is for $f^n=\beta_+^n=0$, when the velocity $v^n$ is expected to experience a single exponential decay. Instead, Eq.~(\ref{eq:GW97_vv}) shows that this is a decay of two exponentials since Eq.~(\ref{eq:GW97_vv}) is a second order difference equation. Further, as the time step is increased, the behavior may even become oscillatory in its decay. The evolution of the velocity $v^n$ is a signature of the corresponding unacceptable behavior of the configurational coordinate $r^n$. Thus, the empirically suggested value in Ref.~\cite{GW97} of $\lambda=\frac{1}{2}$ is, indeed, the {\it only} sensible choice, as $\lambda=\frac{1}{2}$ ensures both that Eq.~(\ref{eq:GW97_rr}) conforms to the form of Eq.~(\ref{eq:Vt_general}), and that Eq.~(\ref{eq:GW97_vv}) for $f^n=\beta_+^n=0$ becomes a first order difference equation in the velocity, supporting, e.g., a single exponential decay of a damped system with no other external forces. We therefore do not consider $\lambda\neq\frac{1}{2}$.\\

\noindent
{\underline{$\lambda=\frac{1}{2}$}}: For this choice of $\lambda$, the integrator of Eqs.~(\ref{eq:GW97_1-3}) and (\ref{eq:GW97_rr}) conforms to Eq.~(\ref{eq:Vt_general}) with the following functional parameters:
\begin{subequations}
\begin{eqnarray}
c_2 & = & 2c_1-1 \; = \; 1-{\gamma\Delta{t}}\label{eq:GW97_c2}\\
c_3 & = & 1 \label{eq:GW97_c3} \\
c_4 & = & 0 \label{eq:GW97_c4} \\
c_5 & = & 0 \label{eq:GW97_c5} \\
c_6 & = & 2 \label{eq:GW97_c6} \\
\zeta & = &0  \,. \label{eq:GW97_zeta} 
\end{eqnarray} \label{eq:GW97_coeff}\noindent
\end{subequations}
Inserting these coefficients into Eqs.~(\ref{eq:DE_discrete}), (\ref{eq:Drift_discrete}), and Eq.~(\ref{eq:Dist_discrete}), yields the results
\begin{subequations}
\begin{eqnarray}
\Gamma_{\rm diff} & = & 1 \label{eq:GW97_G_diff}\\
\Gamma_{\rm drift} & = & 1 \label{eq:GW97_G_drift}\\
\Gamma_{\rm dist} & = & \frac{1}{1-\frac{1}{4c_1}(\Omega_0\Delta{t})^2} \,.\label{eq:GW97_G_dist}
\end{eqnarray}\label{eq:GW97_G}\noindent
\end{subequations}
Notice that the stability criterion is given by a combination of Eqs.~(\ref{eq:Stability_eq_c2}) and (\ref{eq:Stability_eq_R-}), such that
\begin{eqnarray}
\Omega_0\Delta{t} & < & \sqrt{4+\left(\frac{\gamma}{\Omega_0}\right)^2}-\frac{\gamma}{\Omega_0} \,. \label{eq:Stability_GW}
\end{eqnarray}
Figure~\ref{fig:fig_GW97} shows the expressions of Eqs.~(\ref{eq:GW97_G_diff}) and (\ref{eq:GW97_G_drift}) in Fig.~\ref{fig:fig_GW97}a, and Eq.~(\ref{eq:GW97_G_dist}) in Fig.~\ref{fig:fig_GW97}b, the latter for $\gamma=\frac{1}{2}\Omega_0$, $\gamma=\frac{3}{2}\Omega_0$, and $\gamma=4\Omega_0$. The stability limit from Eq.~(\ref{eq:Stability_eq_R-}) is indicated by the vertical arrows in Fig.~\ref{fig:fig_GW97}b. As seen from the figure and Eq.~(\ref{eq:GW97_G}), both normalized diffusion, $\Gamma_{\rm diff}$, and drift velocity, $\Gamma_{\rm drift}$, are correct for all time steps. The behavior of the normalized sampling distribution width, $\Gamma_{\rm dist}$, is given from either Eq.~(\ref{eq:GW97_G_dist}) or (\ref{eq:Dist_discrete}), and is indicating that the configurational temperature is always larger than desired.

\subsection{{{LI$_{02}$}}}
\label{sec:sec_LI}

\begin{figure}[t]
\centering
\scalebox{0.585}{\centering \includegraphics[trim={3.10cm 13.75cm 1cm 5.0cm},clip]{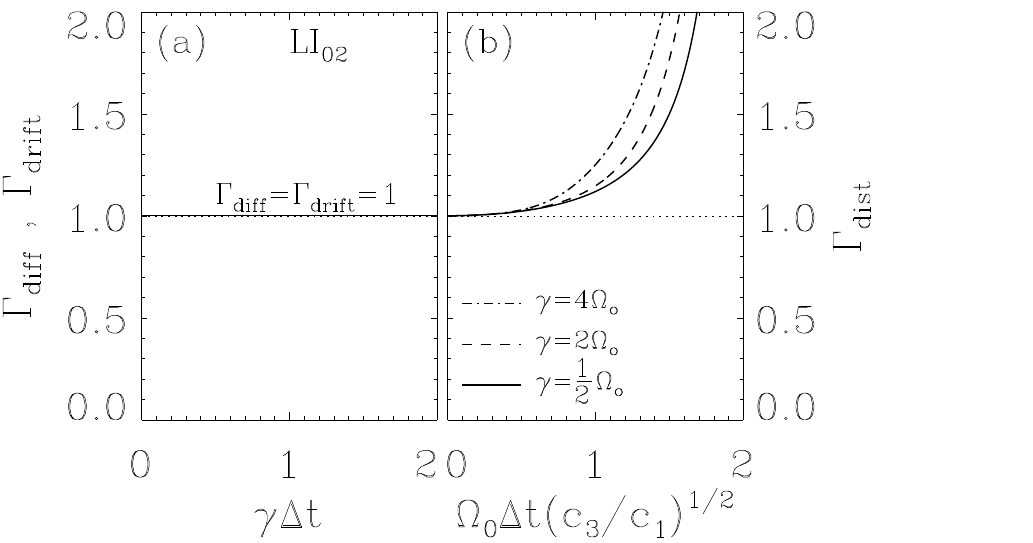}}
\caption{Normalized discrete-time diffusion [$\Gamma_{\rm diff}$, Eqs.~(\ref{eq:DE_discrete}) and (\ref{eq:LI_G_diff})], drift [$\Gamma_{\rm drift}$, Eqs.~(\ref{eq:Drift_discrete}) and (\ref{eq:LI_G_drift})], and Boltzmann configurational temperature [$\Gamma_{\rm dist}$, Eqs.~(\ref{eq:Dist_discrete}) and (\ref{eq:LI_G_dist})] for the LI$_{02}$ integrator of Ref.~\cite{Skeel2002}, as a function (a) of  $\gamma\Delta{t}$ for $\Gamma_{\rm diff}$ (solid) and $\Gamma_{\rm drift}$ (dashed), and (b) of $\Omega_0\Delta{t}$ for $\Gamma_{\rm dist}$, the latter for select values of normalized damping $\gamma/\Omega_0$ as indicated on the figure. The stability range for the time step is given from Eq.~(\ref{eq:Stability_eq_R-}) to be given by $\Omega_0^2\Delta{t}^2<4\frac{c_1}{c_3}$. Discrete-time quantities are normalized to the correct continuous-time quantities, Eq.~(\ref{eq:Basic_truths}).
}
\label{fig:fig_LI}
\end{figure}

The LI$_{02}$ Langevin Impulse integrator \cite{Skeel2002} by Skeel and Izaguirre originates in the configurational form of Eq.~(\ref{eq:Vt_general}) with the functional coefficients
\begin{subequations}
\begin{eqnarray}
c_2 & = & 2c_1-1 \; = \; e^{-\gamma\Delta{t}} \label{eq:LI_c2}\\
c_3 & = & \frac{1-c_2}{{\gamma\Delta{t}}} \; \rightarrow \; 1 \; \; {\rm for} \; \; \gamma\Delta{t}\rightarrow0\label{eq:LI_c3} \\
c_4 & = & 0 \label{eq:LI_c4} \\
c_5 & = & \frac{2}{\gamma\Delta{t}}\sqrt{c_3(1-c_1)-c_2(c_3-c_2)}  \label{eq:LI_c5} \\
&& \rightarrow \; \frac{2}{\sqrt{3}} \; \; {\rm for} \; \; \gamma\Delta{t}\rightarrow0 \nonumber \\
c_6 & = & \frac{2}{\gamma\Delta{t}}\sqrt{1-c_3-c_3(1-c_1)}\label{eq:LI_c6}\\
&& \rightarrow \; \frac{2}{\sqrt{3}} \; \; {\rm for} \; \; \gamma\Delta{t}\rightarrow0 \nonumber  \\
c_5c_6\zeta & = & c_1c_3-c_2 \; \rightarrow \; \frac{1}{2} \; \; {\rm for} \; \; \gamma\Delta{t}\rightarrow0\,. \label{eq:LI_zeta} 
\end{eqnarray} \label{eq:LI_coeff}\noindent
\end{subequations}
Inserting these coefficients into Eqs.~(\ref{eq:DE_discrete}), (\ref{eq:Drift_discrete}), and (\ref{eq:Dist_discrete}) yields the results
\begin{subequations}
\begin{eqnarray}
\Gamma_{\rm diff} & = & 1 \label{eq:LI_G_diff}\\
\Gamma_{\rm drift} & = & 1 \label{eq:LI_G_drift}\\
\Gamma_{\rm dist} & = & \frac{1-\displaystyle\frac{c_1c_3-c_2}{(1-c_2)^2}\frac{c_3}{c_1}(\Omega_0\Delta{t})^2}{1-\displaystyle\frac{c_3}{4c_1}(\Omega_0\Delta{t})^2}  \,. \label{eq:LI_G_dist}
\end{eqnarray}\label{eq:LI_G}\noindent
\end{subequations}
In this case, the two transport characteristics, diffusion and drift, of the three linear benchmark characteristics in Eq.~(\ref{eq:LI_G}), show correct values for all time-steps. The stability criterion of Eq.~(\ref{eq:Stability_eq_R-}) shows that the stability limit is $\sqrt{\frac{c_3}{c_1}}\Omega_0\Delta{t}<2$ (thus, $\Omega_0\Delta{t}<2$ for $\gamma\rightarrow0$), and the stability limit increases as $\gamma$ is increased. This is visible in the width of the sampling distribution represented by $\Gamma_{\rm dist}$ in Eq.~(\ref{eq:LI_G_dist}), which is always larger than the correct value, diverging when $\Omega_0\Delta{t}$ approaches the stability limit, thereby exhibiting elevated configurational temperature in simulations. Figure~\ref{fig:fig_LI} shows the expressions of Eqs.~(\ref{eq:LI_G_diff}) and (\ref{eq:LI_G_drift}) in Fig.~\ref{fig:fig_LI}a and Eq.~(\ref{eq:LI_G_dist}) in Fig.~\ref{fig:fig_LI}b, the latter for $\gamma=\frac{1}{2}\Omega_0$, $\gamma=2\Omega_0$, and $\gamma=4\Omega_0$.

\subsection{RC$_{03}$}
\label{sec:sec_RCA}

\begin{figure}[t]
\centering
\scalebox{0.585}{\centering \includegraphics[trim={3.10cm 13.75cm 1cm 5.0cm},clip]{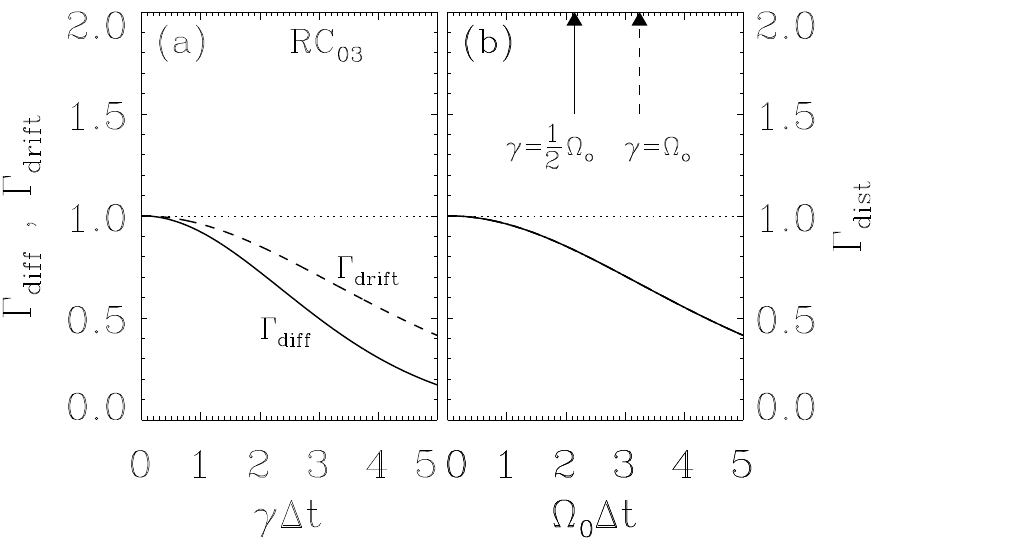}}
\caption{Normalized discrete-time diffusion [$\Gamma_{\rm diff}$, Eqs.~(\ref{eq:DE_discrete}) and (\ref{eq:RC_Im_G_diff})], drift [$\Gamma_{\rm drift}$, Eqs.~(\ref{eq:Drift_discrete}) and (\ref{eq:RC_Im_G_drift})], and Boltzmann configurational temperature [$\Gamma_{\rm dist}$, Eqs.~(\ref{eq:Dist_discrete}) and (\ref{eq:RC_Im_G_dist})] for the RC$_{03}$ integrator of Ref.~\cite{Ricci}, as a function (a) of  $\gamma\Delta{t}$ for $\Gamma_{\rm diff}$ (solid) and $\Gamma_{\rm drift}$ (dashed), and (b) of $\Omega_0\Delta{t}$ for $\Gamma_{\rm dist}$, the latter for any value of normalized damping $\gamma/\Omega_0>0$. Vertical arrows in (b) indicate exemplified stability limits of the time step as given by Eq.~(\ref{eq:Stability_eq_R-}). Discrete-time quantities are normalized to the correct continuous-time quantities, Eq.~(\ref{eq:Basic_truths}).
}
\label{fig:fig_RCA}
\end{figure}

Ricci and Ciccotti suggest in Eqs.~(17) and (18) of Ref.~\cite{Ricci} the following RC$_{03}$ predictor-corrector integrator, which does not generally conform to the Verlet-type form of Eq.~(\ref{eq:Vt_general})
\begin{subequations}
\begin{eqnarray}
\tilde{r}^{n+\frac{1}{2}} & = & r^n+\frac{\Delta{t}}{2}v^n \label{eq:RCA_P}\\
v^{n+1} & = & c_2v^n+\frac{\Delta{t}}{m}\sqrt{c_2}\,f(\tilde{r}^{n+\frac{1}{2}})+\frac{\sqrt{c_2}}{m}\beta_+^n  \label{eq:RCA_v}\\
r^{n+1} & = & \tilde{r}^{n+\frac{1}{2}}+\frac{\Delta{t}}{2}v^{n+1}\,.
\end{eqnarray}
\end{subequations}
Attempting to eliminate the velocity parameter $v^n$, one arrives at
\begin{subequations}
\begin{eqnarray}
\tilde{r}^{n+\frac{1}{2}} & = & r^n+\frac{\Delta{t}}{2}v^n\label{eq:RC_Im_P}\\
r^{n+1} & = & 2c_1r^n-c_2r^{n-1}+\frac{\sqrt{c_2}\Delta{t}^2}{2m}\,\times \label{eq:RC_Im_a}  \\
&&[f(\tilde{r}^{n-\frac{1}{2}})+f(\tilde{r}^{n+\frac{1}{2}})]+\frac{\sqrt{c_2}\Delta{t}}{2m}(\beta_+^{n-1}+\beta_+^{n})\nonumber  \\
r^{n+1}-r^n & = & \frac{\Delta{t}}{2}(v^{n+1}+v^n) \,. \label{eq:RC_Im_2}
\end{eqnarray}\label{eq:RC_Im}\noindent
\end{subequations}
Due to the predictor step in Eqs.~(\ref{eq:RCA_P}) and (\ref{eq:RC_Im_P}) applied to the force evaluations in Eqs.~(\ref{eq:RCA_v}) and Eq.~(\ref{eq:RC_Im_a}), an intermediate velocity parameter, Eq.~(\ref{eq:RC_Im_2}), is necessitated even when attempting the configurational form in Eq.~(\ref{eq:RC_Im}). Thus, Eq.~(\ref{eq:RC_Im_a}) is not a standard configurational Verlet-type method. However, for the linear analysis conducted in this work, linearization of the two force terms in Eq.~(\ref{eq:RC_Im_a}) by a Hooke's force, $f^n=-\kappa r^n$, produces the standard configurational form of Eq.~(\ref{eq:Vt_general}) with the following parameters:
\begin{subequations}
\begin{eqnarray}
c_2 & = & 2c_1-1 \; = \; e^{-{\gamma\Delta{t}}} \label{eq:RC_Im_c2}\\
c_3 & = & \sqrt{c_2} \; \rightarrow \; 1 \; \; {\rm for} \; \; \gamma\Delta{t}\rightarrow0\label{eq:RC_Im_c3_dist}\\
c_4 & = & 0 \label{eq:RC_Im_c4_dist}\\
c_5 & = & \sqrt{c_2}\; \rightarrow \; 1 \; \; {\rm for} \; \; \gamma\Delta{t}\rightarrow0 \label{eq:RC_Im_c5}\\
c_6 & = & \sqrt{c_2} \; \rightarrow \; 1 \; \; {\rm for} \; \; \gamma\Delta{t}\rightarrow0\label{eq:RC_Im_c6}\\
\zeta & = & 1\,. \label{eq:RC_IM_zeta}
\end{eqnarray}\label{eq:RCA_coeff}\noindent
\end{subequations}
Inserting these coefficients into Eqs.~(\ref{eq:DE_discrete}), (\ref{eq:Drift_discrete}), and (\ref{eq:Dist_discrete}) yields the results
\begin{subequations}
\begin{eqnarray}
\Gamma_{\rm diff} & = & \left(\frac{{\gamma\Delta{t}}}{1-c_2}\right)^2c_2\label{eq:RC_Im_G_diff}\\
\Gamma_{\rm drift} & = & \frac{{\gamma\Delta{t}}}{1-c_2}\sqrt{c_2} \label{eq:RC_Im_G_drift}\\
\Gamma_{\rm dist} & = & \frac{{\gamma\Delta{t}}}{1-c_2}\sqrt{c_2}\,.  \label{eq:RC_Im_G_dist}
\end{eqnarray}\label{eq:RC_Im_G}\noindent
\end{subequations}
Notice that, for the calculation of $\Gamma_{\rm drift}$, $f^n={\rm const}$, which implies that $c_3=c_4=\frac{1}{2}\sqrt{c_2}$ instead of the values given for the Hooke's force in Eqs.~(\ref{eq:RC_Im_c3_dist}) and (\ref{eq:RC_Im_c4_dist}). However, the drift result remains the same since $\Gamma_{\rm drift}$ in Eq.~(\ref{eq:Drift_discrete}) depends only on $c_3+c_4$. 

Figure~\ref{fig:fig_RCA} shows the expressions of Eqs.~(\ref{eq:RC_Im_G_diff}) and (\ref{eq:RC_Im_G_drift}) in Fig.~\ref{fig:fig_RCA}a and Eq.~(\ref{eq:RC_Im_G_dist}) in Fig.~\ref{fig:fig_RCA}b, the latter for any value of $\gamma>0$. It follows from Eqs.~(\ref{eq:RC_Im_c2}) and (\ref{eq:Stability_eq_R-}) that a finite stability limit of the normalized time step, $\Omega_0\Delta{t}$, is the same as for the SS$_{78}$ integrator in Sec.~\ref{sec:sec_SS} [see Eqs.~(\ref{eq:SS_stability_espression}) and (\ref{eq:SS_stability}) with associated comments]. The stability limits are exemplified in Fig.~\ref{fig:fig_RCA}b with vertical arrows for $\gamma=\frac{1}{2}\Omega_0$ and $\gamma=\Omega_0$.
As is obvious from both Fig.~\ref{fig:fig_RCA} and Eq.~(\ref{eq:RC_Im_G}), none of the linear benchmark characteristics in Eq.~(\ref{eq:RC_Im_G}) show correct values for $\Omega_0\Delta{t}>0$.

\subsection{VEC$_{06}$}
\label{sec:sec_VECA}

\begin{figure}[t]
\centering
\scalebox{0.585}{\centering \includegraphics[trim={3.10cm 13.75cm 1cm 5.0cm},clip]{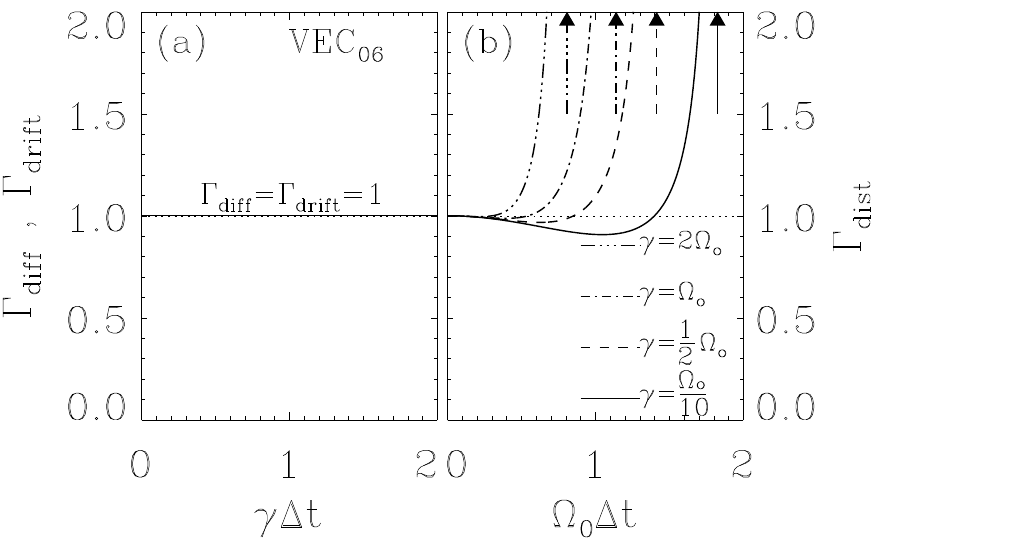}}
\caption{Normalized discrete-time diffusion [$\Gamma_{\rm diff}$, Eqs.~(\ref{eq:DE_discrete}) and (\ref{eq:VE-C_A_Ex_diff})], drift [$\Gamma_{\rm drift}$, Eqs.~(\ref{eq:Drift_discrete}) and (\ref{eq:VE-C_A_Ex_drift})], and Boltzmann configurational temperature [$\Gamma_{\rm dist}$, Eqs.~(\ref{eq:Dist_discrete}) and (\ref{eq:VE-C_A_Ex_dist})] for the VEC$_{06}$ integrator of Ref.~\cite{VEC06}, as a function (a) of  $\gamma\Delta{t}$ for $\Gamma_{\rm diff}$ (solid) and $\Gamma_{\rm drift}$ (dashed), and (b) of $\Omega_0\Delta{t}$ for $\Gamma_{\rm dist}$, the latter for select values of normalized damping $\gamma/\Omega_0$ as indicated on the figure. Vertical arrows in (b) indicate the stability limit of the time step as given by Eq.~(\ref{eq:Stability_eq_R-}). Discrete-time quantities are normalized to the correct continuous-time quantities, Eq.~(\ref{eq:Basic_truths}).
}
\label{fig:fig_VECA}
\end{figure}

With $c_2=1-\gamma\Delta{t}+\frac{1}{2}(\gamma\Delta{t})^2$,  Vanden-Eijnden and Ciccotti proposed in their Eq.~(21) of Ref.~\cite{VEC06} the VEC$_{06}$ integrator,
\begin{subequations}
\begin{eqnarray}
r^{n+1} & = & r^n + \Delta{t}(1-\frac{\Delta{t}}{2})\,v^n+\frac{\Delta{t}}{2m}f^n+\frac{\Delta{t}}{2m}\left[\beta_\xi^n+\frac{1}{\sqrt{3}}\beta_\eta^n\right] \nonumber \\
\\
v^{n+1} & = & c_2v^n+\frac{\Delta{t}}{2m}\left[(1-\gamma\Delta{t})\,f^n+f^{n+1}\right]\nonumber \\
&&+\frac{1}{m}\left[(1-\frac{\gamma\Delta{t}}{2})\beta_\xi^n-\frac{\gamma\Delta{t}}{2\sqrt{3}}\beta_\eta^n\right]\,,
\end{eqnarray}
\end{subequations}
where $\langle\beta_\xi^n\rangle=\langle\beta_\eta^n\rangle=\langle\beta_\xi^n\beta_\eta^\ell\rangle=0$ and $\langle\beta_\xi^n\beta_\xi^\ell\rangle=\langle\beta_\eta^n\beta_\eta^\ell\rangle=2\,\alpha\Delta{t}\,k_BT\,\delta_{n,\ell}$. Notice that the stability criterion given in Eq.~(\ref{eq:Stability_eq_c2}) requires $\gamma\Delta{t}<2$ for this method. The requirement of Eq.~(\ref{eq:Stability_eq_R-}) may limit the time step further. The configurational version of this algorithm is Eq.~(\ref{eq:Vt_general}) with the parameters
\begin{subequations}
\begin{eqnarray}
c_2 & = & 2c_1-1 \; = \; 1-{\gamma\Delta{t}}+\frac{1}{2}({\gamma\Delta{t}})^2 \label{eq:VE-C_A_Ex_c2}\\
c_3 & = &  1-\frac{1}{4}{\gamma\Delta{t}}\; \rightarrow \; 1 \; \; {\rm for} \; \; \gamma\Delta{t}\rightarrow0 \label{eq:VE-C_A_Ex_c3} \\
c_4 & = &-\frac{1}{4}{\gamma\Delta{t}} \; \rightarrow \; 1 \; \; {\rm for} \; \; \gamma\Delta{t}\rightarrow0\label{eq:VE-C_A_Ex_c4} \\
c_5 & = & \sqrt{\frac{4}{3}-2\gamma\Delta{t}+(\gamma\Delta{t})^2} \; \rightarrow \; \frac{2}{\sqrt{3}} \; \; {\rm for} \; \; \gamma\Delta{t}\rightarrow0\nonumber \\
\label{eq:VE-C_A_Ex_c5} \\
c_6 & = &  \frac{2}{\sqrt{3}} \label{eq:VE-C_A_Ex_c6} \\
c_5c_6\zeta & = & \frac{2}{3}-\gamma\Delta{t} \; \rightarrow \; \frac{2}{3} \; \; {\rm for} \; \; \gamma\Delta{t}\rightarrow0 \,. \label{eq:VE-C_A_Ex_zeta}
\end{eqnarray}\label{eq:VE-C_A_coeff}\noindent
\end{subequations}
Inserting these coefficients into Eqs.~(\ref{eq:DE_discrete}), (\ref{eq:Drift_discrete}), and (\ref{eq:Dist_discrete}) yields the results
\begin{subequations}
\begin{eqnarray}
\Gamma_{\rm diff} & = & 1  \label{eq:VE-C_A_Ex_diff}\\
\Gamma_{\rm drift} & = & 1 \label{eq:VE-C_A_Ex_drift}\\
\Gamma_{\rm dist} & = & \displaystyle\frac{1}{1-\frac{\gamma\Delta{t}}{2}+(\frac{\Omega_0\Delta{t}}{2})^2} \,\times \label{eq:VE-C_A_Ex_dist} \\
&& \frac{(1-\frac{\gamma\Delta{t}}{2})^2-\frac{1}{2}\displaystyle\frac{\frac{4}{3}-\frac{5}{3}\gamma\Delta{t}+\frac{(\gamma\Delta{t})^3}{4}}{1-\frac{1}{2}\gamma\Delta{t}+\frac{1}{4}(\gamma\Delta{t})^2}(\frac{\Omega_0\Delta{t}}{2})^2}{(1-\frac{\gamma\Delta{t}}{2})\left[1-\displaystyle\frac{1}{1-\frac{1}{2}\gamma\Delta{t}+\frac{1}{4}(\gamma\Delta{t})^2}(\frac{\Omega_0\Delta{t}}{2})^2\right]} \,. \nonumber
\end{eqnarray}\label{eq:VE-C_A_Ex_G}\noindent
\end{subequations}
The two transport characteristics, diffusion and drift, are correctly mimicked for all stable time steps by this method. The width of the sampling distribution, $\Gamma_{\rm dist}$, given in Eq.~(\ref{eq:VE-C_A_Ex_dist}), is time-step dependent, limiting the correct result for $\Delta{t}\rightarrow0$, as is expected. The behavior of the deviations has two characteristics rooted in the two limitations to the stability range of this method; namely the conditions of Eqs.~(\ref{eq:Stability_eq_R-}) and (\ref{eq:Stability_eq_c2}), the former being the limiting stability condition for small $\gamma/\Omega_0<1$ and the latter for $\gamma/\Omega_0>1$. Figure~\ref{fig:fig_VECA} shows the expressions of Eqs.~(\ref{eq:VE-C_A_Ex_diff}) and (\ref{eq:VE-C_A_Ex_drift}) in Fig.~\ref{fig:fig_VECA}a, and Eq.~(\ref{eq:VE-C_A_Ex_dist}) in Fig.~\ref{fig:fig_VECA}b, the latter for $\gamma=\frac{1}{10}\Omega_0$, $\gamma=\frac{1}{2}\Omega_0$, $\gamma=\Omega_0$, and $\gamma=2\Omega_0$. The stability limits are indicated by the vertical arrows. Notice that at least one of the two conditions always limits stable behavior at or below $\Omega_0\Delta{t}<2$.

\subsection{{{BAOAB$_{12}$}}}
\label{sec:sec_LM}

\begin{figure}[t]
\centering
\scalebox{0.585}{\centering \includegraphics[trim={3.10cm 13.75cm 1cm 5.0cm},clip]{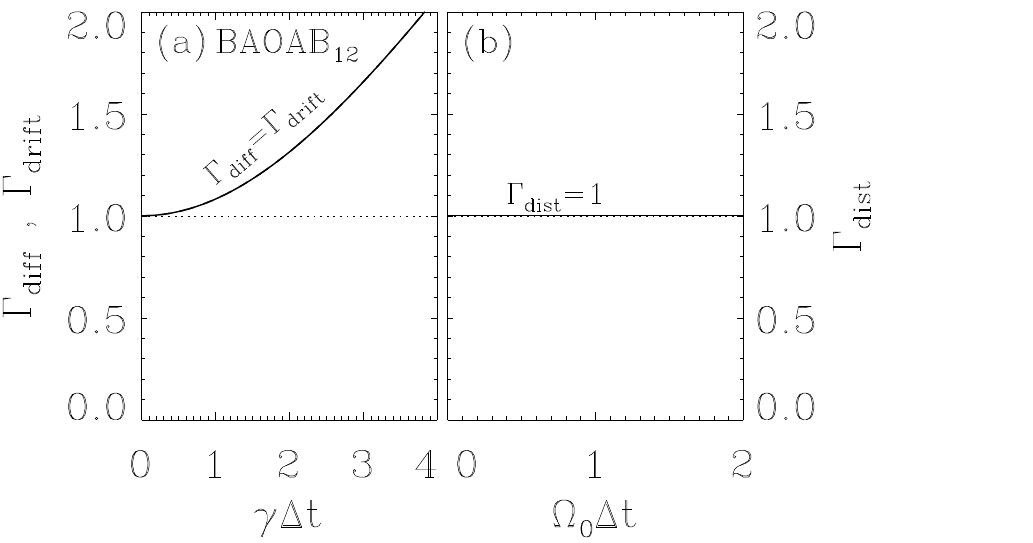}}
\caption{Normalized discrete-time diffusion [$\Gamma_{\rm diff}$, Eqs.~(\ref{eq:DE_discrete}) and (\ref{eq:LM_G_diff})], drift [$\Gamma_{\rm drift}$, Eqs.~(\ref{eq:Drift_discrete}) and (\ref{eq:LM_G_drift})], and Boltzmann configurational temperature [$\Gamma_{\rm dist}$, Eqs.~(\ref{eq:Dist_discrete}) and (\ref{eq:LM_G_dist})] for the BAOAB$_{12}$ integrator of Ref.~\cite{LM}, as a function (a) of  $\gamma\Delta{t}$ for $\Gamma_{\rm diff}$ (solid) and $\Gamma_{\rm drift}$ (dashed), and (b) of $\Omega_0\Delta{t}$ for $\Gamma_{\rm dist}$, the latter for any value of normalized damping $\gamma/\Omega_0>0$. The stability range for the time step is given from Eq.~(\ref{eq:Stability_eq_R-}) to be $\Omega_0\Delta{t}<2$. Discrete-time quantities are normalized to the correct continuous-time quantities, Eq.~(\ref{eq:Basic_truths}).
}
\label{fig:fig_LM}
\end{figure}

With $c_2=\exp(-\gamma\Delta{t})$, the Leimkuhler and Matthews BAOAB$_{12}$ integrator \cite{LM} first appeared in the velocity-Verlet splitting form
\begin{subequations}
\begin{eqnarray}
v^{n+\frac{1}{4}} & = & v^n+\frac{\Delta{t}}{2m}f^n \\
r^{n+\frac{1}{2}} & = & r^n+\frac{\Delta{t}}{2}v^{n+\frac{1}{4}} \\
v^{n+\frac{3}{4}} & = & c_2v^{n+\frac{1}{4}}+\sqrt{\frac{1-c_2^2}{2\gamma\Delta{t}}}\beta_+^n \label{eq:LM_Stoch} \\
r^{n+1} & = & r^{n+\frac{1}{2}}+\frac{\Delta{t}}{2}v^{n+\frac{3}{4}} \\
v^{n+1} & = & v^{n+\frac{3}{4}}+\frac{\Delta{t}}{2m}f^{n+1}\,,
\end{eqnarray}\label{eq:LM_orig}\noindent
\end{subequations}
and the corresponding configurational algorithm, Eq.~(\ref{eq:Vt_general}), is given by the following functional parameters:
\begin{subequations}
\begin{eqnarray}
c_2 &  = &  2c_1-1 \; = \; e^{-{\gamma\Delta{t}}} \label{eq:LM_c2}\\
c_3 & = & c_1\; \rightarrow \; 1 \; \; {\rm for} \; \; \gamma\Delta{t}\rightarrow0 \label{eq:LM_c3} \\
c_4 & = & 0 \label{eq:LM_c4} \\
c_5 & = & \sqrt{c_1\frac{1-c_2}{{\gamma\Delta{t}}}} \; \rightarrow \; 1 \; \; {\rm for} \; \; \gamma\Delta{t}\rightarrow0\label{eq:LM_c5} \\
c_6 & = & \sqrt{c_1\frac{1-c_2}{{\gamma\Delta{t}}}} \; \rightarrow \; 1 \; \; {\rm for} \; \; \gamma\Delta{t}\rightarrow0\label{eq:LM_c6} \\
\zeta & = & 1\,.  \label{eq:LM_zeta} 
\end{eqnarray} \label{eq:LM_coeff}\noindent
\end{subequations}
As pointed out in Ref.~\cite{Kieninger}, the GROMACS molecular modeling suite \cite{Gromacs} is conducting the same configurational sampling as BAOAB$_{12}$.
Inserting the coefficients of Eq.~(\ref{eq:LM_coeff}) into Eqs.~(\ref{eq:DE_discrete}), (\ref{eq:Drift_discrete}), and (\ref{eq:Dist_discrete}) yields the results
\begin{subequations}
\begin{eqnarray}
\Gamma_{\rm diff} & = & c_1\frac{{\gamma\Delta{t}}}{1-c_2} \label{eq:LM_G_diff}\\
\Gamma_{\rm drift} & = & c_1\frac{{\gamma\Delta{t}}}{1-c_2} \label{eq:LM_G_drift}\\
\Gamma_{\rm dist} & = & 1\,. \label{eq:LM_G_dist}
\end{eqnarray}\label{eq:LM_G}\noindent
\end{subequations}
The two transport characteristics, diffusion and drift, are incorrectly mimicked by this integrator as seen from Eqs.~(\ref{eq:LM_G_diff}) and (\ref{eq:LM_G_drift}) as well as Fig.~\ref{fig:fig_LM}a. In contrast, the width of the sampling distribution, $\Gamma_{\rm dist}$, given in Eq.~(\ref{eq:LM_G_dist}) and shown in Fig.~\ref{fig:fig_LM}b, demonstrates the first example of an integrator that can precisely reproduce the Boltzmann distribution in discrete time. The stability range of this integrator is $\Omega_0\Delta{t}<2$, and is given by Eq.~(\ref{eq:Stability_eq_R-}).

\subsection{{{GJ$_{13-20}$}}}
\label{sec:sec_GJ}

\begin{figure}[t]
\centering
\scalebox{0.585}{\centering \includegraphics[trim={3.10cm 13.75cm 1cm 5.0cm},clip]{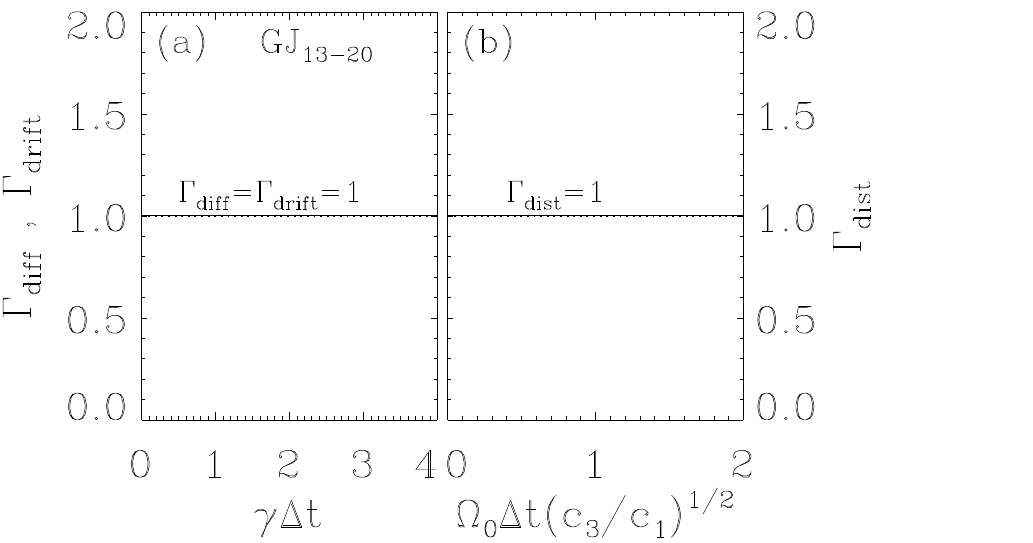}}
\caption{Normalized discrete-time diffusion [$\Gamma_{\rm diff}$, Eqs.~(\ref{eq:DE_discrete}) and (\ref{eq:GJ_G_diff})], drift [$\Gamma_{\rm drift}$, Eqs.~(\ref{eq:Drift_discrete}) and (\ref{eq:GJ_G_drift})], and Boltzmann configurational temperature [$\Gamma_{\rm dist}$, Eqs.~(\ref{eq:Dist_discrete}) and (\ref{eq:GJ_G_dist})] for the GJ$_{13-20}$ integrator of Ref.~\cite{GJ}, as a function (a) of  $\gamma\Delta{t}$ for $\Gamma_{\rm diff}$ (solid) and $\Gamma_{\rm drift}$ (dashed), and (b) of $\Omega_0\Delta{t}$ for $\Gamma_{\rm dist}$, the latter for any value of normalized damping $\gamma/\Omega_0>0$. The stability range for the time step is $\Omega_0^2\Delta{t}^2<4\frac{c_1}{c_3}$ given from Eq.~(\ref{eq:Stability_eq_R-}). Discrete-time quantities are normalized to the correct continuous-time quantities, Eq.~(\ref{eq:Basic_truths}).
}
\label{fig:fig_GJ}
\end{figure}

For any functional parameter, $c_2$ with the limiting behavior $c_2\rightarrow1$$-$$\gamma\Delta{t}$ for $\gamma\Delta{t}$$\rightarrow$$0$, the complete GJ$_{13-20}$ integrator set \cite{GJ} was derived in the configurational form of Eq.~(\ref{eq:Vt_general}), specifically for correct simulations of the three linear characteristics, diffusion, drift, and configurational sampling such that $\Gamma_{\rm diff}=1$ [from  Eq.~(\ref{eq:DE_discrete})], $\Gamma_{\rm drift}=1$ [from Eq.~(\ref{eq:Drift_discrete})], and $\Gamma_{\rm dist}=1$ [from Eq.~(\ref{eq:Dist_discrete})] for any time-step within the stability range. The set of integrators is characterized by the functional coefficients
\begin{subequations}
\begin{eqnarray}
c_2 & = & 2c_1-1  \; \rightarrow \;  1-\gamma\Delta{t} \; \;  \; {\rm for} \; \gamma\Delta{t}\rightarrow0 \label{eq:_GJ_c2} \\
c_3 & = & \frac{1-c_2}{{\gamma\Delta{t}}} \; \rightarrow \; 1 \; \; {\rm for} \; \; \gamma\Delta{t}\rightarrow0\label{eq:_GJ_c3} \\
c_4 & = & 0 \label{eq:_GJ_c4} \\
c_5 & = & c_3 \; \rightarrow \; 1 \; \; {\rm for} \; \; \gamma\Delta{t}\rightarrow0\label{eq:_GJ_c5} \\
c_6 & = & c_3 \; \rightarrow \; 1 \; \; {\rm for} \; \; \gamma\Delta{t}\rightarrow0\label{eq:_GJ_c6} \\
\zeta & = & 1\,, \label{eq:_GJ_zeta} 
\end{eqnarray} \label{eq:GJ_coeff}\noindent
\end{subequations}
and, by design, the resulting normalized linear benchmark quantities are:
\begin{subequations}
\begin{eqnarray}
\Gamma_{\rm diff} & = & 1 \label{eq:GJ_G_diff}\\
\Gamma_{\rm drift} & = & 1 \label{eq:GJ_G_drift}\\
\Gamma_{\rm dist} & = & 1\,. \label{eq:GJ_G_dist}
\end{eqnarray}\label{eq:GJ_G}\noindent
\end{subequations}
Thus, this set of methods, only differentiated by the choice of the coefficient $c_2$, yields both correct transport properties and correct Boltzmann sampling for any reasonable one-time-step velocity attenuation parameter, $|c_2|<1$, with the limiting behavior, $c_2\rightarrow1-{\gamma\Delta{t}}$ for ${\gamma\Delta{t}}\rightarrow0$. The stability range for the time step is given by Eq.~(\ref{eq:Stability_eq_R-}) to be $\Omega_0\Delta{t}<2\sqrt{c_1/c_3}$. Figure~\ref{fig:fig_GJ} displays the trivial GJ$_{13-20}$ benchmarks as a visual reference to other integrators.

This set of integrators was argued in Ref.~\cite{GJ} to capture all stochastic Verlet-type integrators that can reproduce all three linear benchmarks correctly, using $c_4=0$ and only a single stochastic variable per time step, $\zeta=1$. The statement was strengthened in Ref.~\cite{Josh_2020} by indicating that the three sought-after features in Eq.~(\ref{eq:GJ_G}) {\it require} $c_4=0$ and $\zeta=1$. In conjunction with the general expressions for $\Gamma_{\rm diff}$ [Eqs.~(\ref{eq:DE_discrete})], $\Gamma_{\rm drift}$ [Eq.~(\ref{eq:Drift_discrete})], and $\Gamma_{\rm dist}$ [Eq.~(\ref{eq:Dist_discrete})], derived in Appendices \ref{app:app_Diff}, \ref{app:app_Drift}, and \ref{app:app_Dist}, Appendix~\ref{app:app_Exclusive} concisely solidifies that {\it any} stochastic Verlet-type integrator satisfying Eq.~(\ref{eq:GJ_G}) must be described by Eq.~(\ref{eq:GJ_coeff}).

The first identified method of the exclusive GJ$_{13-20}$ set is the GJF integrator \cite{GJF1}, originally derived in the standard velocity-Verlet form
\begin{subequations}
\begin{eqnarray}
r^{n+1} & = & r^n+c_1\left[\Delta{t}\,v^n+\frac{\Delta{t}^2}{2m}f^n+\frac{\Delta{t}}{2m}\beta_+^n\right] \\
v^{n+1} & = & c_2v^n+\frac{\Delta{t}}{2m}(c_2f^n+f^{n+1})+\frac{c_1}{m}\beta_+^n
\end{eqnarray}
with\begin{eqnarray}
c_2 & = & \frac{1-\frac{1}{2}\gamma\Delta{t}}{1+\frac{1}{2}\gamma\Delta{t}} \,, \label{eq:GJF_c2}
\end{eqnarray}\noindent
\end{subequations}
such that $c_3=c_1$. A leap-frog version of this algorithm was later introduced with a half-step velocity that produces the precise kinetic temperature \cite{2GJ}, and this was generalized to all GJ$_{13-20}$ methods in Ref.~\cite{GJ}. 
As seen from Eq.~(\ref{eq:GJF_c2}) the GJF/GJ-I method appears with the $c_2$ parameter coinciding with that of the BBK$_{84}$ integrator, Eq.~(\ref{eq:BBK_c2}). It is noticeable that replacing the exponential form of $c_2$ in the BAOAB$_{12}$ integrator with $c_2$ from Eq.~(\ref{eq:BBK_c2}) or (\ref{eq:GJF_c2}) yields the GJF/GJ-I integrator with all the properties of Eq.~(\ref{eq:GJ_G}). Thus, the GJF/GJ-I integrator can be written as Eq.~(\ref{eq:LM_orig}) with $c_2$ from Eq.~(\ref{eq:BBK_c2}); see also Refs.~\cite{Gauss_noise,GJ24}. By scaling the time step in the BAOAB$_{12}$ integrator, it was later shown in Ref.~\cite{Sivak} that such revision leads to an integrator also of the configurational GJ form given in Eq.~(\ref{eq:GJ_G}), this one with $c_2$ given by Eq.~(\ref{eq:LM_c2}). See also Ref.~\cite{Josh_3} for a discussion of the relationships between the GJ$_{13-20}$ and ABO splitting methods.

Many applicable $c_2$ parameters can be proposed, each defining a GJ$_{13-20}$ integrator. Apart from the two already mentioned, four more are suggested in Ref.~\cite{GJ}, one in Ref.~\cite{Josh_3}, and one more in Ref.~\cite{GJ24}. Additionally, any one of the $c_2$ parameters of the methods analyzed in this paper can be used to define a GJ$_{13-20}$ integrator. In fact, the GW$_{97}$ $c_2$ parameter from Eq.~(\ref{eq:GW97_c2}) can be used in Eq.~(\ref{eq:GJ_coeff}) to obtain the GJ-III integrator proposed in Ref.~\cite{GJ}. A particular feature of this method is that $c_3=c_5=c_6=1$.

Several GJ$_{13-20}$ integrators, including the original GJ-I integrator, GJF \cite{GJF1} augmented by the statistically robust half-step velocity \cite{2GJ}, are implemented in 
 the molecular modeling suite LAMMPS \cite{Plimpton2,Linke} as well as in the RUMD suite \cite{RUMD}.

\begin{table}[t]
\centering
\scalebox{0.9}{\centering{
\begin{tabular}{|c||c|c|c|}\hline
                & \makebox[0.95 in][c]{Zero force} & \makebox[0.95 in][c]{Constant force} & \makebox[0.95 in][c]{Hooke's force} \\  \hline
                & Einstein & Terminal & Configurational \\ 
Integrator & diffusion &  drift velocity & temperature\\ 
                & $\Gamma_{\rm diff}$ $\left[k_BT/\alpha\right]$ & $\Gamma_{\rm drift}$  $\left[f/\alpha\right]$ & $\Gamma_{\rm dist}$  $\left[T\right]$ \\ \hline \hline 
SS$_{\rm 78}$ & 1+${\cal O}[(\gamma\Delta{t})^2]$ & 1+${\cal O}[(\gamma\Delta{t})^2]$ & 1+${\cal O}[(\Omega_0\Delta{t})^2]$ \\ \hline
EB$_{\rm 80}$ & 1 & 1 & 1+${\cal O}[(\Omega_0\Delta{t})^1]$ \\ \hline
MPA$_{\rm 80-82}$ & 1 & 1+${\cal O}[(\gamma\Delta{t})^2]$ & 1+${\cal O}[(\Omega_0\Delta{t})^2]$ \\ \hline
vGB$_{\rm 82}$ & 1+${\cal O}[(\gamma\Delta{t})^1]$ & 1 & 1+${\cal O}[(\Omega_0\Delta{t})^1]$ \\ \hline
BBK$_{\rm 84}$ & 1& 1 & 1+${\cal O}[(\Omega_0\Delta{t})^2]$ \\ \hline
vGB$_{\rm 88}$ & 1 & 1+${\cal O}[(\gamma\Delta{t})^2]$ & 1+${\cal O}[(\Omega_0\Delta{t})^2]$ \\ \hline
GW$_{\rm 97}$ & 1 & 1 & 1+${\cal O}[(\Omega_0\Delta{t})^2]$ \\ \hline
LI$_{\rm 02}$ & 1 & 1 & 1+${\cal O}[(\Omega_0\Delta{t})^2]$ \\ \hline
RC$_{\rm 03}$ & 1+${\cal O}[(\gamma\Delta{t})^2]$ & 1+${\cal O}[(\gamma\Delta{t})^2]$ & 1+${\cal O}[(\Omega_0\Delta{t})^2]$ \\ \hline
VEC$_{\rm 06}$ & 1 & 1 & 1+${\cal O}[(\Omega_0\Delta{t})^2]$ \\ \hline
BAOAB$_{\rm 12}$ & 1+${\cal O}[(\gamma\Delta{t})^2]$ & 1+${\cal O}[(\gamma\Delta{t})^2]$ & 1 \\ \hline
GJ$_{\rm 13-20}$ & 1 & 1 & 1 \\ \hline
\end{tabular}
}}
\caption{Normalized discrete-time diffusion [$\Gamma_{\rm diff}$, from Eq.~(\ref{eq:DE_discrete})], drift [$\Gamma_{\rm drift}$, from Eq.~(\ref{eq:Drift_discrete})], and Boltzmann configurational temperature [$\Gamma_{\rm dist}$, from Eq.~(\ref{eq:Dist_discrete})] for the specific integrators considered in this work. Relative to the characteristic units, given in Eq.~(\ref{eq:Basic_truths}), unity indicates correct algorithm response, and the limiting time-step deviations from perfection are indicated by their ${\cal O}(\cdot)$ scaling. For detailed time-step dependence of these three fundamental measures, see Figs.~\ref{fig:fig_SS}-\ref{fig:fig_GJ} and their respective Sections~\ref{sec:sec_SS}-\ref{sec:sec_GJ}.
}
\label{tbl:tbl_1}
\end{table}

\section{Discussion}
\label{sec:Discussion}
We have provided a general framework for analyzing the quality of stochastic Verlet-type integrators through the three most basic properties of linear systems; namely the two transport measures, a) diffusion on a flat surface and b) drift on a tilted planar surface, as well as c) the statistical sampling of the configurational space in a harmonic potential. Only configurational measures are considered in this work since many different velocity definitions can be tailored to any given configurational method, and we therefore consider velocity analysis a separate exercise, which should be conducted once a desirable configurational algorithm has been selected (see Ref.~\cite{GJ24}). We have exemplified, in chronological order, the use of the derived general expressions for the three characteristic linear measures through application to twelve representative integrators that have been considered over the past almost five decades, and we have visualized the resulting measures as a function of the two reduced time steps, $\gamma\Delta{t}$ and $\Omega_0\Delta{t}$, to illuminate what one should expect from simulations. These results, coarsely summarized in Table~\ref{tbl:tbl_1}, which outlines the scaling of the statistical errors of interest as a function of the time step $\Delta{t}$, have been verified through implementation and direct numerical simulations of the presented algorithms.

Due to its wide-spread historical use in molecular dynamics, we here briefly comment on the Nos{\'e}-Hoover extended Hamiltonian thermostat \cite{Hoover_book,Nose,Hoover,Holian}, even if it is not covered by the analysis of this work. Because of the inherent necessity of a velocity coordinate to calibrate the thermal energy in this algorithm, and since this deterministic approach uses either of the second order on-site or half-step velocity measures given in Refs.~\cite{Swope,Beeman,Buneman,Hockney}, this thermostat exhibits second order errors in, e.g., the configurational temperature, which will typically be elevated due to the properties of central difference velocity-response to convex potentials (see appendix in Ref.~\cite{2GJ}).

We point out that, as solidified in Appendix~\ref{app:app_Exclusive} based on the general expressions in this work, {\it only} the tightly connected set of GJ$_{13-20}$ integrators can exactly reproduce all three characteristic measures for any time step within the stability criteria. This set of integrators has remarkably simple algorithmic coefficients, demands that only a single stochastic variable per time step is used, and has no requirement for memory to store a previous-time force. Thus, given that all stochastic Verlet-type integrators can generally be written in the form of Eq.~(\ref{eq:Vt_general}), and therefore all will induce similar computational load per time step, the statistical benefits of the GJ methods come with neither additional computational cost nor complexity regarding implementation. We here point back to the penultimate paragraph of the Introduction that outlines the importance of securing accuracy for the basic statistical measures in linear systems before applying an algorithm to a nonlinear and complex system, for which a simulation cannot be reasonably expected to fare better than for the corresponding linear system. This notion is also supported by comparative simulation results given in, e.g., Refs.~\cite{GJ,2GJ,Finkelstein_1,Josh_2020,Agarwal_2025}, all demonstrating high statistical accuracy within the GJ framework, even for rather large time steps, when simulating nonlinear and complex systems.
Further, high-quality velocity definitions, with optimal statistical accuracy of kinetic measures, as well as practical and efficient implementations, have previously been published for the GJ methods \cite{GJ24}.

Finally, we restate that two of the integrators investigated here, BBK$_{84}$ and GJ$_{13-20}$, are implemented, well tested, and available in LAMMPS \cite{Plimpton2,Linke,SSvsBBK}. GJ$_{13-20}$ integrators are also implemented and tested in RUMD \cite{RUMD}.

\section{Acknowledgments}
The author acknowledges useful discussions with Bogdan Tanygin and Simone Melchionna at the outset of this work as well as Giovanni Ciccotti for pointing the author to the VEC$_{06}$ integrator. The author also appreciates ongoing insightful conversations with Joshua Finkelstein.\\

\section{Data Availability Statement}
All data presented in this work are results of the derived closed expressions.

\section{Funding and/or Competing Interests}
No funding was received for conducting this study.
The author has no relevant financial or non-financial interests to disclose.

\appendix
\section{Discrete-Time Diffusion}
\label{app:app_Diff}
The most basic demonstration of diffusion is that of a flat potential surface with friction that is balanced by thermal fluctuations. This potential surface implies $f^n=0$ in Eq.~(\ref{eq:Vt_general}). From Eq.~(\ref{eq:Diff_rr}) the diffusion of the calculation of the discrete-time configurational coordinate is
\begin{eqnarray}
D_E & = & \lim_{n\rightarrow\infty}\frac{\left\langle(r^{n}-r^0)^2\right\rangle}{2\,n\Delta{t}}\,, \label{eq:D_E_flat}
\end{eqnarray}
where
\begin{eqnarray}
r^n-r^0 & = & \Delta{t}\sum_{\ell=0}^{n-1}\frac{r^{\ell+1}-r^{\ell}}{\Delta{t}} \,. \label{eq:rnmr0}
\end{eqnarray}
From Eq.~(\ref{eq:Vt_general}) (with $f^n=0$) we find
\begin{eqnarray}
r^\ell-r^{\ell-1} & = & c_2^\ell(r^{1}-r^0)+\frac{\Delta{t}}{2m}\sum_{k=0}^{\ell-1}c_2^k(c_5\beta_-^{\ell-k}+c_6\beta_+^{\ell-k}) \,, \nonumber \\ \label{eq:rnmr0_recur}
\end{eqnarray}
which, when inserted into Eq.~(\ref{eq:rnmr0}) for $n\ge1$ and $|c_2|<1$, gives
\begin{eqnarray}
r^n-r^0 & = &  \frac{1-c_2^n}{1-c_2}(r^1-r^0)\nonumber \\
&&+\frac{\Delta{t}}{2m}\sum_{q=1}^{n-1}\frac{1-c_2^{n-q}}{1-c_2}(c_5\beta_-^q+c_6\beta_+^q)\,,
\end{eqnarray}
where $q=\ell-k$. Using Eq.~(\ref{eq:d_noise_all}) we then get
\begin{eqnarray}
&&\left\langle(r^n-r^0)^2\right\rangle \; = \; \left(\frac{1-c_2^n}{1-c_2}\right)^2(r^1-r^0)^2\\
&&+{2\,\alpha\Delta{t}\,k_BT}\left(\frac{\Delta{t}}{2m}\right)^2\frac{c_5^2+c_6^2}{(1-c_2)^2}\sum_{q=1}^{n-1}\left(1-c_2^{n-q}\right)^2 \nonumber \\
&& + {2\,\alpha\Delta{t}\,k_BT} \left(\frac{\Delta{t}}{2m}\right)^2\frac{2c_5c_6\zeta}{(1-c_2)^2}\,\sum_{q=1}^{n-2}(1-c_2^{n-1-q})(1-c_2^{n-q})\,.\nonumber 
\end{eqnarray}
Equation~(\ref{eq:D_E_flat}) then gives
\begin{subequations}
\begin{eqnarray}
D_E & = &\frac{k_BT}{\alpha}\,\Gamma_{\rm diff}  \label{eq:D_E_flat_result}\\
\Gamma_{\rm diff} & = & \left(\frac{{\gamma\Delta{t}}}{1-c_2}\right)^2 \frac{c_5^2+c_6^2+2c_5c_6\zeta}{4} \,, \label{eq:Gamma_diff}
\end{eqnarray}\label{eq:DE_discrete}\noindent
\end{subequations}
where $\Gamma_{\rm diff}$ is the normalized diffusion constant. Thus, any integrator of the form given in Eq.~(\ref{eq:Vt_general}) that is required to reproduce the correct diffusion, Eq.~(\ref{eq:Diff_rr}),  on a flat energy surface must yield $\Gamma_{\rm diff}=1$.

\section{Discrete-Time Drift}
\label{app:app_Drift}
The most basic drift property is the system response to a tilted planar potential surface with friction. This potential surface implies $f^n=f={\rm const}$. From Eq.~(\ref{eq:Drift_r}) the configurational calculation of the discrete-time drift velocity is given by
\begin{eqnarray}
v_d & = & \frac{\left\langle r^{n+1}-r^n\right\rangle}{\Delta{t}} \,,\label{eq:drift_wd}
\end{eqnarray}
which from Eqs.~(\ref{eq:Vt_general}) and (\ref{eq:d_noise_ave}) directly reads
\begin{subequations}
\begin{eqnarray}
v_d & = & \frac{f}{\alpha}\, \Gamma_{\rm drift} \label{eq:drift_wd_result}\\
\Gamma_{\rm drift} & = & \frac{{\gamma\Delta{t}}}{1-c_2}(c_3+c_4) \,, \label{eq:Gamma_drift}
\end{eqnarray}\label{eq:Drift_discrete}\noindent
\end{subequations}
where $\Gamma_{\rm drift}$ is the normalized drift velocity. Thus, any integrator of the form given in Eq.~(\ref{eq:Vt_general}) that is required to reproduce the correct drift velocity, Eq.~(\ref{eq:Drift_r}), on a tilted planar energy surface must yield $\Gamma_{\rm drift}=1$.

\section{Discrete-Time Boltzmann Distribution}
\label{app:app_Dist}
The harmonic potential surface, $E_p(r^n)=\frac{1}{2}\kappa r^n$, yields the linear Hooke's force, $f^n=-\kappa r^n=-m\Omega_0^2r^n$, where $\Omega_0$ is the natural frequency of the corresponding harmonic oscillator without damping.
Following Ref.~\cite{Josh_2020}, we write the linearized Eq.~(\ref{eq:Vt_general}) in the form
\begin{eqnarray}
r^{n+1} & = & 2c_1Xr^n-c_2Yr^{n-1}+\frac{\Delta{t}}{2m}(c_5\beta_-^{n}+c_6\beta_+^n)\,,\; \label{eq:Vt_harmonic}
\end{eqnarray}
with
\begin{subequations}	
\begin{eqnarray}
X & = & 1-\frac{c_3}{c_1}\frac{(\Omega_0\Delta{t})^2}{2} \label{eq:XX}\\
Y & = & 1+\frac{c_4}{c_2}(\Omega_0\Delta{t})^2 \,. \label{eq:YY}
\end{eqnarray}\label{eq:XXYY}\noindent
\end{subequations}
Multiplying Eq.~(\ref{eq:Vt_harmonic}) with $r^{n-1}$, $r^n$, and $r^{n+1}$, we can write the equations for the three moments, $\langle r^{n-1}r^{n+1}\rangle$, $\langle r^nr^{n+1}\rangle$, and $\langle r^nr^n\rangle$, as follows
\begin{eqnarray}
&&\left(\begin{array}{ccc}
1 & -2c_1X & c_2Y \\
0 & 1+c_2Y & -2c_1X \\
c_2Y & -2c_1X & 1\end{array}\right)
\left(\begin{array}{c}
\langle r^{n-1}r^{n+1}\rangle \\
\langle r^nr^{n+1}\rangle\\
\langle r^nr^n\rangle\end{array}\right) \nonumber \\
&& = \;
\frac{\Delta{t}}{2m}\left(\begin{array}{c} c_5\cancel{\langle r^{n-1}\beta_-^n\rangle}+c_6\cancel{\langle r^{n-1}\beta_+^n\rangle} \\
c_5{\langle r^{n}\beta_-^n\rangle}+c_6\cancel{\langle r^{n}\beta_+^n\rangle} \\ 
c_5{\langle r^{n+1}\beta_-^n\rangle}+c_6{\langle r^{n+1}\beta_+^n\rangle} \end{array}\right)\,,  \label{eq:matrix_cor}
\end{eqnarray}
where causality dictates that the stroked terms be zero, and where, from Eq.~(\ref{eq:Vt_harmonic}), 
\begin{subequations}
\begin{eqnarray}
\langle r^n\beta_-^n\rangle & = & \frac{\Delta{t}}{2m}c_6\langle\beta_+^{n-1}\beta_-^n\rangle \\
\langle r^{n+1}\beta_+^n\rangle & = & \frac{\Delta{t}}{2m}c_6\langle\beta_+^{n}\beta_+^n\rangle \\
\langle r^{n+1}\beta_-^n\rangle & = & 2c_1Xc_5\langle r^n\beta_-^n\rangle+\frac{\Delta{t}}{2m}c_6\langle\beta_+^{n-1}\beta_-^n\rangle\,.
\end{eqnarray}\noindent
\end{subequations}
The crucial second moment, $\langle r^nr^n\rangle$ from Eq.~(\ref{eq:Corr_rr}), can now be extracted in discrete time from Eq.~(\ref{eq:matrix_cor}):
\begin{subequations}
\begin{eqnarray}
&&\left\langle r^nr^n\right\rangle \; = \; \frac{k_BT}{\kappa}\,\Gamma_{\rm dist}\label{eq:rnrn_result} \\
&&\Gamma_{\rm dist} \; = \; \frac{{\gamma\Delta{t}}}{1-c_2-c_4(\Omega_0\Delta{t})^2} \,\times \label{eq:Gamma_dist} \\
&&\frac{2c_1\left[c_5^2+c_6^2+2c_5c_6\zeta\right]+\left[c_4(c_5^2+c_6^2)-2c_3c_5c_6\zeta\right](\Omega_0\Delta{t})^2}{2(c_3+c_4)[4c_1-(c_3-c_4)\,(\Omega_0\Delta{t})^2]} ,\nonumber
\end{eqnarray}\label{eq:Dist_discrete}\noindent
\end{subequations}
where $\Gamma_{\rm dist}$ is the normalized variance of the sampling distribution, and, equivalently, $\Gamma_{\rm dist}$ is also the normalized configurational temperature. Thus, any integrator of the form given in Eq.~(\ref{eq:Vt_general}) that is required to reproduce the correct Boltzmann distribution, and equivalently the configurational temperature, Eq.~(\ref{eq:Corr_rr}), for a simple harmonic potential must yield $\Gamma_{\rm dist}=1$.

As pointed out in Sec.~\ref{sec:Specific_Int}, unlike $\Gamma_{\rm diff}$ and $\Gamma_{\rm drift}$,  $\Gamma_{\rm dist}$ is a function of two different reduced time steps, $\gamma\Delta{t}$ and $\Omega_0\Delta{t}$, the latter dependency being explicitly represented in Eq.~(\ref{eq:Gamma_dist}). Given that the numerator of Eq.~(\ref{eq:Gamma_dist}) is a polynomial of degree no higher than two in $\Omega_0\Delta{t}$, while the denominator is a polynomial of degree no higher than four in $\Omega_0\Delta{t}$, we see that a necessary condition for simulating the correct Boltzmann distribution requires that either $c_4=0$ or $c_4=c_3$. These two cases are explored in Appendix \ref{app:app_Exclusive}.

\section{Integrator Stability}
\label{app:app_Stability}

As noted previously, stability requires that the one-time-step velocity attenuation parameter, $c_2$ in Eq.~(\ref{eq:Vt_general}), is shorter than unity for $\gamma\Delta{t}>0$:
\begin{eqnarray}
|c_2| & < 1\,. \label{eq:Stability_eq_c2}
\end{eqnarray}
Additionally, stability is analyzed from applying the harmonic potential to the damped, noiseless system to see if the solution is contracting or diverging. Thus, the relevant equation is
\begin{eqnarray}
r^{n+1} & = & 2c_1Xr^n-c_2Yr^{n-1} \,, \label{eq:Stability_eq_0}
\end{eqnarray}
from where we obtain the characteristic eigenvalues,
\begin{eqnarray}
\Lambda_{\pm} & = & c_1X\pm\sqrt{c_1^2X^2-c_2Y} \,, \label{eq:Stability_eq_L}
\end{eqnarray}
such that the solution to Eq.~(\ref{eq:Stability_eq_0}) is given by a linear combination of $r^n\sim\Lambda_\pm^n$ (notice that the superscript $n$ on $\Lambda_\pm$ is an exponent). Stability is therefore implied by $|\Lambda_\pm|\le1$.
\begin{description}
\item[{\underline{Complex $\Lambda_\pm$: $c_2Y>c_1^2X^2$.}}] From Eq.~(\ref{eq:Stability_eq_L}), the length of the eigenvalues is given by
\begin{eqnarray}
|\Lambda_\pm|^2 & = & c_2Y\,, \label{eq:Stability_eq_C_length}
\end{eqnarray}
and stability is given by
\begin{eqnarray}
c_4\,(\Omega_0\Delta{t})^2 & < 1-c_2 \,. \label{eq:Stability_eq_C}
\end{eqnarray}
Notice that, for complex $\Lambda_\pm$ and $|c_2|<1$, stability is ensured for any reduced time step, $\Omega_0\Delta{t}$, if $c_4\le0$.

In this underdamped regime we can write the eigenvalues of Eq.~(\ref{eq:Stability_eq_L}) as
\begin{eqnarray}
\Lambda_\pm & = & \sqrt{c_2Y}\exp(\pm i \, \Omega_V\Delta{t})\,, \label{eq:Stability_L_exp}
\end{eqnarray}
where the frequency, $\Omega_V$, is that of the damped discrete-time oscillation. Thus, the frequency can be determined from
\begin{subequations}
\begin{eqnarray}
&&\cos{\Omega_V\Delta{t}} \; = \; \frac{c_1-\frac{c_3}{2}(\Omega_0\Delta{t})^2}{\sqrt{c_2+c_4(\Omega_0\Delta{t})^2}} \label{eq:Stability_eq_OV_cos}\\
&&\sin{\Omega_V\Delta{t}} \; = \; \Omega_0\Delta{t} \label{eq:Stability_eq_OV_sin} \,\times \\
&&\frac{\sqrt{c_1c_3+c_4-c_3^2\left(\frac{\Omega_0\Delta{t}}{2}\right)^2-\left(\frac{1-c_2}{{\gamma\Delta{t}}}\right)^2\left(\frac{\alpha}{2m\Omega_0}\right)^2}}{\sqrt{c_2+c_4(\Omega_0\Delta{t})^2}} \,. \nonumber
\end{eqnarray}\label{eq_Stability_OV}\noindent
\end{subequations}

\item[{\underline{Real $\Lambda_\pm$: $c_2Y<c_1^2X^2$.}}] Stability is here given from Eqs.~(\ref{eq:XXYY}) and (\ref{eq:Stability_eq_L}):
\begin{subequations}
\begin{eqnarray}
|\Lambda_+| & < & 1 \; \; \Rightarrow \; \; c_3+c_4 \; > \; 0 \label{eq:Stability_eq_R+}\\
|\Lambda_-| & < & 1 \; \; \Rightarrow \; \; (c_3-c_4)\,(\Omega_0\Delta{t})^2 \; < \; 4c_1 \,,\label{eq:Stability_eq_R-}
\end{eqnarray}\label{eq:Stability_eq_R}\noindent
\end{subequations}
where Eq.~(\ref{eq:Stability_eq_R+}) is self-evident, e.g., from a meaningful drift velocity; see Eq.~(\ref{eq:Drift_discrete}) or Eq.~(\ref{eq:Vt_general}). It then follows that an instability for Real $\Lambda_\pm$, where $\Lambda_-<\Lambda_+$, occurs when $\Lambda_-<-1$, which is the case when Eq.~(\ref{eq:Stability_eq_R-}) is invalid.
\end{description}
Regardless of the eigenvalues $\Lambda_\pm$ being complex or real, the stability range is given by $|\Lambda_\pm|<1$. Therefore, the discrete-time evolution of the simulated system in one time step is given by $\Lambda_\pm$, and not directly by the time step $\Delta{t}$. It follows that the stability range of $\Delta{t}$ itself is not a good measure of how far one can evolve the system over that one time step, since a large stability range in $\Delta{t}$ is only sustained by the method by slowing down the motion compared to a method with smaller stability range in $\Delta{t}$. Thus, all methods with similar statistical properties, regardless of the stability range in $\Delta{t}$, will sample phase space with similar efficiency when operated close to their respective stability limits.

\section{The Exclusivity of the GJ Set of Integrators}
\label{app:app_Exclusive}

It was argued in Refs.~\cite{GJ} that this set of integrators is complete in the sense that no other algorithm of the form given in Eq.~(\ref{eq:Vt_general}) can accomplish all three characteristic properties given in Eq.~(\ref{eq:GJ_G}) and outlined in detail in Appendices \ref{app:app_Diff}, \ref{app:app_Drift}, and \ref{app:app_Dist}. This assertion was subsequently strengthened in Ref.~\cite{Josh_2020}. Based on the general expressions derived in the mentioned Appendicies, we here concisely and methodically solidify that only the GJ set of integrators correctly reproduces the three most basic linear properties.

As stated at the end of Appendix~\ref{app:app_Dist}, in reference to Eq.~(\ref{eq:Dist_discrete}), a necessary condition for correct configurational sampling distribution, $\Gamma_{\rm dist}=1$, is that either $c_4=c_3$ or $c_4=0$. 

\subsection{The case $c_4=c_3$}
The expression for $\Gamma_{\rm dist}$ in Eq.~(\ref{eq:Dist_discrete}) becomes
\begin{eqnarray}
&&\Gamma_{\rm dist} \; = \; \gamma\Delta{t}\,\times \\
&& \frac{2c_1\left[c_5^2+c_6^2+2c_5c_6\zeta\right]+c_3\left[c_5^2+c_6^2-2c_5c_6\zeta\right](\Omega_0\Delta{t})^2}{16c_1c_3\left[1-c_2-c_3(\Omega_0\Delta{t})^2\right]}\,.\nonumber 
\end{eqnarray}
The condition $\Gamma_{\rm dist}=1$ yields the two conditions
\begin{subequations}
\begin{eqnarray}
\gamma\Delta{t}\left[c_5^2+c_6^2+2c_5c_6\zeta\right] & = & 8c_3(1-c_2) \label{eq:GJ_cond_1a}\\
\gamma\Delta{t}\left[c_5^2+c_6^2-2c_5c_6\zeta\right] & = & -16c_1c_3\,.  \label{eq:GJ_cond_1b}
\end{eqnarray}\label{eq:GJ_cond_1m}\noindent
\end{subequations}
Adding and subtracting the two equations in Eq.~(\ref{eq:GJ_cond_1m}) yield
\begin{subequations}
\begin{eqnarray}
\gamma\Delta{t}\left[c_5^2+c_6^2\right] & = & 4c_3(1-c_2)-8c_1c_3 \label{eq:GJ_cond_1c}\\
\gamma\Delta{t}\left[c_5c_6\zeta\right] & = & 2c_3(1-c_2)+4c_1c_3\,, \label{eq:GJ_cond_1d}
\end{eqnarray}\label{eq:GJ_cond_1}\noindent
\end{subequations}
where Eq.~(\ref{eq:GJ_cond_1c}) implies that
\begin{eqnarray}
0 & < & 1-c_2-2c_1 \; = \; -2c_2\label{eq:c4_cond_fail}
\end{eqnarray}
with $2c_1=1+c_2$ from Eq.~(\ref{eq:c1}), emphasizing that $0<c_1<1$ for $|c_2|<1$, and with $c_4+c_3>0$, which is necessary for a meaningful method (see, e.g., Eq.~(\ref{eq:Drift_discrete})). Given that the one-time-step attenuation parameter $c_2\rightarrow1$ for $\gamma\Delta{t}\rightarrow0$, Eq.~(\ref{eq:c4_cond_fail}) shows that no meaningful stochastic integrator can be constructed to yield $\Gamma_{\rm dist}=1$ for $c_4=c_3$.

\subsection{The case $c_4=0$}

We immediately notice that the condition, $\Gamma_{\rm drift}=1$, from Eq.~(\ref{eq:Drift_discrete}) yields
\begin{eqnarray}
c_3 & = & \frac{1-c_2}{\gamma\Delta{t}} \,. \label{eq:GJ_c3_final}
\end{eqnarray}

The expression for $\Gamma_{\rm dist}$ in Eq.~(\ref{eq:Dist_discrete}) becomes
\begin{eqnarray}
\Gamma_{\rm dist} & = & \frac{{\gamma\Delta{t}}}{1-c_2} \frac{2c_1\left[c_5^2+c_6^2+2c_5c_6\zeta\right]-2c_3c_5c_6\zeta(\Omega_0\Delta{t})^2}{2c_3\left[4c_1-c_3(\Omega_0\Delta{t})^2\right]}.\nonumber \\
\end{eqnarray}
The condition $\Gamma_{\rm dist}=1$ yields the two conditions
\begin{subequations}
\begin{eqnarray}
c_5^2+c_6^2+2c_5c_6\zeta\; = \; & 4&c_3\frac{1-c_2}{\gamma\Delta{t}} \label{eq:GJ_cond_2a}\\
c_5c_6\zeta\; =\; & &c_3\frac{1-c_2}{\gamma\Delta{t}} \,. \label{eq:GJ_cond_2b}
\end{eqnarray}\label{eq:c4zero_A}\noindent
\end{subequations}
We notice from Eq.~(\ref{eq:GJ_cond_2b}) that $\zeta=0$ will not produce a reasonable set of coefficients, given that both $c_3$ and $(1-c_2)$ must be positive (see, e.g., Eqs.~(\ref{eq:Stability_eq_c2}) and (\ref{eq:Drift_discrete})). Thus, $\zeta\neq0$. Equation (\ref{eq:c4zero_A}) can also be written
\begin{subequations}
\begin{eqnarray}
c_5^2+c_6^2 \; = \; & &2c_3\frac{1-c_2}{{\gamma\Delta{t}}} \label{eq:GJ_cond_2c}\\
(c_5^2-c_6^2)^2 \; = \; & - &4\left(c_3\frac{1-c_2}{{\gamma\Delta{t}}}\right)^2\frac{1-\zeta^2}{\zeta^2} \,. \label{eq:GJ_cond_2d}
\end{eqnarray}\label{eq:GJ_cond_2}\noindent
\end{subequations}\noindent
Equation (\ref{eq:GJ_cond_2d}) can only be satisfied for
\begin{eqnarray}
\zeta^2 & = & 1 \label{eq:GJ_zeta2_final}\\
c_5^2 & = & c_6^2 \,, \label{eq:GJ_c52c62_final}
\end{eqnarray}
for which Eq.~(\ref{eq:GJ_cond_2c}) then dictates that
\begin{eqnarray}
c_5^2 & = & c_6^2 \; = \; c_3\frac{1-c_2}{{\gamma\Delta{t}}} \; = \; c_3^2\,,
\end{eqnarray}
the latter equality resulting from Eq.~(\ref{eq:GJ_c3_final}). Finally, Eq.~(\ref{eq:GJ_cond_2b}) specifies that
\begin{eqnarray}
c_5 & = & \zeta c_6\,, \label{eq:GJ_c5c6zeta_final}
\end{eqnarray}
with $\zeta=\pm1$, indicating that the binary choice of $\zeta$ is inconsequential.

With Eqs.~(\ref{eq:GJ_c52c62_final}) and (\ref{eq:GJ_c5c6zeta_final}), the correct discrete-time diffusion, characterized by $\Gamma_{\rm diff}=1$ in Eq.~(\ref{eq:DE_discrete}), implies
\begin{eqnarray}
c_5^2 & = & c_6^2 \; = \; \left(\frac{1-c_2}{\gamma\Delta{t}}\right)^2 \; = \; c_3^2\,,
\end{eqnarray}
where the last equality again is due to $\Gamma_{\rm drift}$ from Eq.~(\ref{eq:GJ_c3_final}).

In summary, unless $c_4=0$, $c_5=\zeta c_6=c_3$, $\zeta^2=1$, and $c_3=\frac{1-c_2}{\gamma\Delta{t}}$ with $c_2\rightarrow1-\gamma\Delta{t}$ for $\gamma\Delta{t}\rightarrow0$, at least one of the linear characteristic measures of Eqs.~(\ref{eq:DE_discrete}), (\ref{eq:Drift_discrete}), and (\ref{eq:Dist_discrete}) is imperfect.
Thus, the GJ set of parameters given in Eq.~(\ref{eq:GJ_coeff}) is the {\it only} possibility for stochastic Verlet-type integrators to exactly reproduce all three basic statistical properties, diffusion, drift, and configurational sampling indicated in Eq.~(\ref{eq:GJ_G}) for all time steps within the stability limit.

 

\begin{thebibliography}{99} 
\bibitem{Toxsvaerd} S.~Toxsvaerd, {\it Newton$^\prime$s discrete dynamics}, Eur.\ Phys.\ J.\ Plus {\bf 135}, 267 (2020).
\bibitem{Stormer_1921} For a review, see, e.g., E.~Hairer, C.~Lubich, and G.~Wanner, {\it Geometric numerical integration illustrated by the St{\o}rmer-Verlet method}, Acta Numerica {\bf 12}, 399 (2003).
\bibitem{Verlet} L. Verlet, {\it Computer experiments on classical fluids. {I}. Thermodynamical properties of Lennard-Jones molecules}, Phys. Rev. {\bf 159}, 98 (1967).
\bibitem{Swope} W.~C.~Swope, H.~C.~Andersen, P.~H.~Berens, K.~R.~Wilson, {\it A computer simulation method for the calculation of equilibrium constants for the formation of physical clusters of molecules: Application to small water clusters}, J.\ Chem.\ Phys.\ {\bf 76}, 637 (1982).
\bibitem{Beeman} D.~Beeman, {\it Some multistep methods for use in molecular dynamics calculations}, J.\ Comp.\ Phys.\ {\bf 20}, 130 (1976).
\bibitem{Buneman} O.~Buneman, {\it Time-reversible difference procedures}, J.\ Comp.\ Phys.\ {\bf 1}, 517 (1967).
\bibitem{Hockney} R.~W.~Hockney, {\it The potential calculation and some applications}, Methods Comput. Phys.\ {\bf 9}, 135 (1970).
\bibitem{Tuckerman} M.~Tuckerman, B.~J.~Berne and G.~J.~Martyna, {\it Reversible multiple time scale molecular dynamics}, J.\ Chem.\ Phys.\ {\bf 97}, 1990 (1992).

\bibitem{AllenTildesley} M.~P.~Allen, D.~J.~Tildesley, {\it Computer Simulation of Liquids}, (Oxford University Press, Inc., 1989).
\bibitem{Frenkel} D.~Frenkel and B.~Smit, {\it Understanding Molecular Simulations: From Algorithms to Applications}, (Academic Press, San Diego, 2002).
\bibitem{Rapaport} D.~C.~Rapaport, {\it The Art of Molecular Dynamics Simulations}, (Cambridge University Press, Cambridge, 2004).
\bibitem{Hoover_book} W.~M.~Hoover, {\it Computational Statistical Mechanics}, (Elsevier Science B.V., 1991).
\bibitem{Leach}A.~Leach,  {\it Molecular Modeling: Principles and Applications}, 2nd ed., (Prentice Hall: Harlow, England, 2001).

\bibitem{SS} T.~Schneider and E.~Stoll, {\it Molecular-dynamics study of a three-dimensional one-component model for distortive phase transitions}, Phys. Rev. B {\bf 17}, 1302 (1978).
\bibitem{Ermak1980}D.~L.~Ermak and H.~Buckholtz, {\it Numerical integration of the Langevin equation: Monte Carlo simulation}, J.~Comp.~Phys.~{\bf 35}, 169 (1980).
\bibitem{Allen80} M.~P.~Allen, {\it Brownian Dynamics Simulation of a Chemical Reaction in Solution}, Mol.\ Phys.\ {\bf 40}, 1073 (1980).
\bibitem{Allen82} M.~P.~Allen, {\it Algorithms for Brownian Dynamics}, Mol.\ Phys.\ {\bf 47}, 599 (1982).
\bibitem{Thalmann2007}F.~Thalmann and J.~Farago, {\it Trotter Derivation of Algorithms for Brownian and Dissipative Particle Dynamics}, J.~Chem.~Phys.~{\bf 127}, 124109 (2007).
\bibitem{vGB82}W.~F.~F. van Gunsteren and H.~J.~C.~Berendsen, {\it Algorithms for Brownian dynamics}, Mol.~Phys. {\bf 45}, 637 (1982).
\bibitem{BBK} A. Br\"{u}nger, C. L. Brooks, and M. Karplus, {\it Stochastic boundary conditions for molecular dynamics simulations of ST2 water}, Chem. Phys. Lett. {\bf 105}, 495 (1984).
\bibitem{vGB88}W.~F.~F. van Gunsteren and H.~J.~C.~Berendsen, {\it A leap-frog algorithm for stochastic dynamics}, Mol.~Sim. {\bf 1}, 173 (1988).
\bibitem{GW97}R.~D.~Groot and P.~B.~Warren, {\it Dissipative Particle Dynamics: Bridging the Gap Between Atomistic and Mesoscopic Simulation}, J.~Chem.~Phys.~{\bf 107}, 4423 (1997).
\bibitem{Skeel2002}R.~D.~Skeel and J.~A.~Izaguirre, {\it An Impulse Integrator for Langevin Dynamics}, Mol.~Phys.~{\bf 100}, 3885 (2002).
\bibitem{Ricci} A.~Ricci and G.~Ciccotti, {\it Algorithms for Brownian Dynamics}, Mol.\ Phys.\ {\bf 101}, 1927 (2003).
\bibitem{VEC06}E.~Vanden-Eijnden and G.~Ciccotti, {\it Second-order integrators for Langevin equations with holonomic constraints}, Chem.~Phys.~Lett.~{\bf 429}, 310 (2006).
\bibitem{LM} B.~Leimkuhler and C.~Matthews, {\it Rational Construction of Stochastic Numerical Methods for Molecular Sampling}, Appl. Math. Res. Express {\bf 2013}, 34 (2012).
\bibitem{GJ}N.~Gr{\o}nbech-Jensen, {\it Complete set of stochastic Verlet-type thermostats for correct Langevin simulations}, Mol.\ Phys.\ {\bf 118}, e1662506 (2020).

\bibitem{Pastor_88} R.~W.~Pastor, B.~R.~Brooks, and A.~Szabo, {\it An analysis of the accuracy of Langevin and molecular dynamics algorithms}, Mol.\ Phys.\ {\bf 65}, 1409 (1988).
\bibitem{2GJ} L.~F.~G.~Jensen and N.~Gr{\o}nbech-Jensen,  {\it Accurate configurational and kinetic statistics in discrete-time Langevin systems}, Mol.\ Phys.\ {\bf 117}, 2511 (2019).
\bibitem{Finkelstein_1}J.~Finkelstein, G.~Fiorin, and S.~Seibold, {\it Comparison of modern Langevin integrators for simulations of coarse-grained polymer melts}, Mol. Phys. {\bf 118}, e1649493 (2020).
\bibitem{Tanygin2024}B.~Tanygin and S.~Melchionna, {\it Comparison of Effective and Stable Langevin Dynamics Integrators}, Comput.~Phys.~Commun.~{\bf 299}, 109152 (2024).

\bibitem{GJ24}N.~Gr{\o}nbech-Jensen, {\it On the Definition of Velocity in Discrete-Time, Stochastic Langevin Simulations}, J.~Stat.~Phys.~{\bf 191}, 137 (2024).

\bibitem{Langevin} P.~Langevin, {\it On the Theory of Brownian Motion}, C.~R.~Acad.~Sci.~Paris {\bf 146}, 530 (1908).
\bibitem{Langevin_Eq} W.~T.~Coffey and Y.~P.~Kalmykov, {\it The Langevin Equation}, 3rd ed., World Scientific, Singapore, 2012.
\bibitem{Parisi} G.~Parisi, {\it Statistical Field Theory}, (Addison-Wesley, Menlo Park, 1988).

\bibitem{Landau} L.\ D.\ Landau and E.\ M.\ Lifshitz, {\it Statistical Physics, Part I} (Pergamon, Oxford, 1980).
\bibitem{Rugh} H.~H.~Rugh, {\it Dynamical Approach to Temperature}, Phys.\ Rev.\ Lett.\ {\bf 78}, 772 (1997).
\bibitem{Powles_05} J.~G.~Powles, G.~Rickayzen, and D.~M.~Heyes, {\it Temperaturs: old, new and middle aged}, Mol.\ Phys.\ {\bf 103}, 1361 (2005).
\bibitem{Saw_23} S.~Saw, L.~Costigliola, and J.~Dyre, {\it Configurational temperature in active matter. I. Lines of invariant physics in the phase diagram of the Ornstein-Uhlenbeck model}, Phys.\ Rev.\ E {\bf 107}, 024609 (2023).

\bibitem{Gauss_noise} N.~Gr{\o}nbech-Jensen, {\it On the Application of Non-Gaussian Noise in Stochastic Langevin Simulations}, J.\ Stat.\ Phys.\ {\bf 190}, 96 (2023).
\bibitem{Josh_2020}J.\ Finkelstein, C.\ Cheng, G.\ Fiorin, B.\ Seibold, and N.\ Gr{\o}nbech-Jensen, {\it The Challenge of Stochastic St{\o}rmer-Verlet Thermostats Generating Correct Statistics}, J.\ Chem.\ Phys.\ {\bf 153}, 134101 (2020).

\bibitem{Plimpton2} A.~P.~Thompson, H.~M.~Aktulga, R.~Berger, D.~S.~Bolintineanu, W.~M.~Brown,
P.~S.~Crozier, P.~J.~in~$^\prime$t Veld, A.~Kohlmeyer, S.~G.~Moore, T.~D.~Nguyen, R.~Shan,
M.~Stevens, J.~Tranchida, C.~Trott, S.~J.~Plimpton, {\it LAMMPS -- a flexible simulation tool for particle-based materials modeling at the atomic, meso, and continuum scales}, Compute.\ Phys.\ Commun. {\bf 271}, 108171 (2022).

\bibitem{Espresso_4.2.0} F.~Weik, R.~Weeber, K.~Szuttor, K.~Breitsprecher, J.~de~Graaf, M.~Kuron, J.~Landsgesell, H.~Menke, D.~Sean and C.~Holm, {\it ESPResSo 4.0 -- an extensible software package for simulating soft matter systems}, Euro. Phys.~J.~Special Topics {\bf 227}, 1789 (2019). See also https://espressomd.github.io/ documentation for ESPResSo-4.2.0 Sec.~6.3.1.

\bibitem{Kieninger} S.~Kieninger and B.~G.~Keller, {\it GROMACS Stochastic Dynamics and BAOAB Are Equivalent Configurational Sampling Algorithms}, J.~ Chem.~Theo.~Comput.\ {\bf 18}, 5792 (2022).

\bibitem{Gromacs} M.~J.~Abraham, T.~Murtola, R.~Schulz, S.~P{\'a}ll, J.~C.~Smith, B.~Hess, and E.~Lindahl, {\it Gromacs: High performance molecular simulations through multi-level parallelism from laptops to supercomputers}, SoftwareX {\bf 1}, 19 (2015).

\bibitem{GJF1} N.~Gr{\o}nbech-Jensen and~O. Farago, {\it A simple and effective Verlet-type algorithm for simulating Langevin dynamics}, Mol. Phys. {\bf 111}, 983 (2013).
\bibitem{Sivak} D.~A.~Sivak, J.~D.~Chodera, and G.~E.~Crooks, {\it Time Step Rescaling Recovers Continuous-Time Dynamical Properties for Discrete-Time Langevin Integration of Nonequilibrium Systems}, J.\ Phys.\ Chem.\ B {\bf 118}, 6466 (2014).
\bibitem{Josh_3}J.~Finkelstein, C.~Cheng, G.~Fiorin, B.~Seibold, and N.~Gr{\o}nbech-Jensen, {\it Bringing Langevin splitting methods into agreement with correct discrete-time thermodynamics}, J.\ Chem.\ Phys.\ {\bf 155}, 184104 (2021).

\bibitem{Linke} GJ implementation in LAMMPS \cite{Plimpton2}; Tim Linke, the LAMMPS team, and this author, unpublished. See ``fix~gjf" in documentation:  https://docs.lammps.org/latest/fix\_gjf.html. Previously, the GJ-I/GJF method was linked through ``fix~langevin" in LAMMPS.

\bibitem{RUMD} N.~P.~Bailey, T.~S.~Ingebrigtsen, J.~Schmidt~Hansen, A.~A.~Veldhorst, L.~B{\o}hling, C.~A.~Lemarchand, A.~E.~Olsen, A.~K.~Bacher, L.~Costigliola, U.~R.~Pedersen, H.~Larsen, J.~C.~ Dyre, T.~B.~Schr{\o}der, {\it RUMD: A general purpose molecular dynamics package optimized to utilize GPU hardware down to a few thousand particles},
SciPost Physics {\bf 3}, 038 (2017). See also rumd.org, Roskilde University Molecular Dynamics package.

\bibitem{SSvsBBK} The implemented default stochastic integrator in LAMMPS \cite{Plimpton2} with ``fix~langevin" has, until recently, been attributed to SS$_{78}$ \cite{SS} in the manual. However, the actual implementation has been, and still is, BBK$_{84}$ \cite{BBK} with the standard on-site central difference velocity of Refs.~\cite{Swope,Beeman}.

\bibitem{Nose} S.~Nos{\'e}, {\it Structure, melting and transport properties of binary liquid pd-si metal alloys: Molecular dynamics simulations}, J.~Chem.~Phys.~{\bf 81}, 511 (1984).

\bibitem{Hoover} W.~G.~Hoover, {\it Canonical dynamics: Equilibrium phase-space distributions}, Phys.~Rev.~A~{\bf 31}, 1695 (1985).

\bibitem{Holian} W.~G.~Hoover and B.~L.~Holian, {\it Kinetic moments method for the canonical ensemble distribution}, Phys.~Lett.~A~{\bf 211}, 253 (1996).

\bibitem{Agarwal_2025}S.~Agarwal, S.~V.~Sukhomlinov, M.~Honecker, and M.~H.~M{\"u}ser, {\it  Advanced Langevin thermostats: Properties, extensions to rheology, and a lean momentum-conserving approach}, J.~Chem.~Phys.\ {\bf 163}, 124110 (2025).
\end{thebibliography}
\end{document}